\documentstyle[12pt]{article}
\begin{document}
\setcounter{section}{0}
\begin{center}
\Large\bf The masses and decay widths of heavy hybrid mesons 
\end{center}
\vspace{0.5cm}
\begin{center}
{ Shi-Lin Zhu}\\\vspace{3mm}
{Institute of Theoretical Physics \\
Academia Sinica, P.O.Box 2735\\
Beijing 100080, China\\
FAX: 086-10-62562587\\
TEL: 086-10-62541816\\
E-MAIL: zhusl@itp.ac.cn}
\end{center}
\vspace{1.0cm}
\begin{abstract}
We first derive the mass sum rules for the
heavy hybrid mesons to obtain the binding energy and 
decay constants in the leading order of
heavy quark effective theory. The pionic couplings between the lightest 
$1^{-+}$ hybrid $(Q\bar q g)$ and the lowest three heavy meson 
doublets are calculated with the light cone QCD sum rules. 
With $SU_f(3)$ flavor symmetry we calculate the widths for 
all the possible two-body decay processes with a Goldstone 
boson in the final state. The total width of the $1^{-+}$ hybrid 
is estimated to be around $300$ MeV. We find that the dominant decay mode 
of the $1^{-+}$ heavy hybrid is $1^{-+} \to \pi + 1^+ $ where the $1^+$ 
heavy meson belongs to the $(1^+, 2^+)$ heavy meson 
doublet. Its branching ratio is about $80\%$ so this mode can be used 
for the experimental search of the lowest heavy hybrid meson. 
\end{abstract}



\pagenumbering{arabic}

\section{ Introduction}
\label{sec1} 

From quark model we know that 
a $q{\bar q}$ meson with orbital angular momentum l and total spin s must have 
$P=(-1)^{l+1}$ and $C=(-1)^{l+s}$. Thus a resonance with $J^{PC}=0^{--}, 
0^{+-}, 1^{-+}, 2^{+-}, \cdots$ must be exotic. Such a state could be a 
gluonic excitation such as hybrids, glueballs or multiquark states.
The hybrid and glueball has been a missing link in the hadron spectrum.
Recently there appears experimental evidence for a $J^{PC}=1^{-+}$ exotic
\cite{e852,cb,ves}. The emergence of evidence for hybrids 
indicates the presence of dynamical glue in QCD and will be a direct 
test of low energy sector of QCD.

The light $1^{-+}$ hybrid meson mass has been found to lie between 
$1\mbox{GeV} \sim 1.7$GeV \cite{balitsky}-\cite{deviron} with QCD 
sum rules \cite{svz}. With the same method the hybrid meson 
containing one heavy quark and heavy hybrid quarkonium 
has been studied in \cite{deviron-heavy}. 
Some decay modes of the light $1^{-+}$ hybrid meson are discussed 
in \cite{deviron-decay}. Recently the masses of the hybrid 
quarkonium $(Q\bar Q g)$ states are calculated in the
limit of $m_Q\to\infty$ \cite{zhu-bbg}.

The hybrid meson spectrum have been studied extensively with other 
theoretical approaches including the constituent gluon models \cite{gluon-mass},
the flux tube models \cite{flux-mass}, the MIT bag model \cite{mit}, 
and the lattice gauge theory \cite{lattice}. Their decays have been
studied in the constituent gluon model via constituent gluon 
dissociation \cite{gluon-decay}, in the different versions of the 
flux tube model with a $^3P_0$ pair creation mechanism \cite{flux}. 
The constituent gluon model predicts a significant width for the 
$\eta \pi$ channel for the light $1^{-+}$ hybrid meson \cite{gluon-decay}
while the width of this channel is very small in other models 
\cite{deviron-decay,flux}. In contrast the flux tube model predicts
very characteristic decay modes for the $1^{-+}$ decays. The dominant
modes are $1^{-+}\to b_1\pi, f_1 \pi$ or $a_1\pi$. In this model the width
of the $\rho\pi$ channel is also significant. 

The QCD sum rule was used to calculate the $\rho\pi, \eta\pi$ channels of 
the $1^{-+}$ hybrid mesons with the three point correlation functions at 
the symmetric point \cite{deviron-decay}. The final sum rule for the 
decay coupling constants suffers from large continuum and excited states 
contamination. 

Since most of the decay calculation has focused on the light hybrid mesons,  
it will prove valuable to calculate the mass and decays of the 
hybrid meson with a heavy quark to see whether the same characteristic
decay modes exist as in the flux tube model.

The combination of the heavy quark effective theory (HQET) \cite{hqet}
and the light cone QCD sum rules \cite{bely-89} provides 
a convenient framework to calculate the strong and electromagnetic 
decays of heavy hadrons containing one heavy quark 
\cite{bely-95,zhu-pion,zhu-radiat}.

In this work we first derive the mass sum rules for the heavy hybrid 
mesons and perform the numerical analysis. Then we calculate the 
pionic couplings between the lightest $1^{-+}$ hybrid $(Q\bar q g)$ 
and the lowest three heavy meson doublets.
Invoking $SU_f(3)$ flavor symmetry for the coupling constants 
we calculate the widths for all the possible two-body strong decay 
processes with a Goldstone boson in the final state. 

\section{The mass sum rules for the heavy hybrid mesons in HQET}
\label{sec2}
\setcounter{equation}{0}
\subsection{Heavy quark effective theory}

The effective Lagrangian of the HQET, up to order $1/m_Q$, is
\begin{equation}
\label{Leff}
   {\cal L}_{\rm eff} = \bar h_v\,i v\!\cdot\!D\,h_v
   + \frac{1}{2 m_Q}\,{\cal K}
   + \frac{1}{2 m_Q}\,{\cal S}+{\cal O}(1/m_Q^2) \,,
\end{equation}
where $h_v(x)$ is the velocity-dependent field related to the
original heavy-quark field $Q(x)$ by
\begin{equation}
   h_v(x) = e^{i m_Q v\cdot x}\,\frac{1+\rlap/v}{2}\,Q(x)\;,
\end{equation}
$v_\mu$ is the heavy hadron velocity. ${\cal K}$ is the kinetic operator defined as
\begin{equation}
\label{kinetic}
{\cal K}=\bar h_v\,(i D_t)^2 h_v\;,
\end{equation}
where $D_t^\mu=D^\mu-(v\cdot D)\,v^\mu$, with 
$D^\mu=\partial^\mu-i g\, A^\mu$ is the gauge-covariant derivative, 
and $\cal S$ is the chromomagnetic operator
\begin{equation}
\label{pauli}
{\cal S}=\frac{g}{2}\,C_{mag}(m_Q/\mu)\;
   \bar h_v\,\sigma_{\mu\nu} G^{\mu\nu} h_v\;,
\end{equation}
where $C_{mag}=\displaystyle{\left(\alpha_s(m_Q)\over
\alpha_s(\mu)\right)^{3/{\beta}_0}}$, ${\beta}_0=11-2n_f/3$.

\subsection{The derivation of the sum rules}

The interpolating current for the $J^{PC}=1^{-+}, 0^{++}$ heavy 
hybrid mesons in HQET reads
\begin{equation}
\label{cu1}
J_\mu (x) ={\bar q}(x)g_s \gamma^\nu G^a_{\mu\nu}(x) {\lambda^a\over 2}
 h_v (x)\;,
\end{equation}
and for the $J^{PC}=1^{+-}, 0^{--}$ ones
\begin{equation}
\label{cu2}
J^5_\mu (x) ={\bar q}(x)g_s \gamma^\nu\gamma_5 G^a_{\mu\nu}(x) {\lambda^a\over 2}
 h_v (x)\;.
\end{equation}

We consider the correlators 
\begin{equation}\label{co-1}
i\int d^4 x e^{ikx} \langle 0| T\{J_\mu (x), J^{\dag}_\nu (0)\} |0\rangle  
=-(g_{\mu\nu} -v_\mu v_\nu )\Pi_1 (\omega)
+v_\mu v_\nu \Pi_2 (\omega) \;,
\end{equation}
\begin{equation}\label{co-2}
i\int d^4 x e^{ikx} \langle 0| T\{J^5_\mu (x), J^{5 \dag}_\nu (0)\} |0\rangle  
=-(g_{\mu\nu} -v_\mu v_\nu )\Pi_3 (\omega)
+v_\mu v_\nu \Pi_4 (\omega) 
\end{equation}
with $\omega =k\cdot v$. The imaginary parts of $\Pi_1, \Pi_2, \Pi_3, \Pi_4$
receive contributions from the $J^{PC}=1^{-+}, 0^{++}, 1^{+-}, 0^{--}$ 
hybrid intermediate states respectively.

To simplify the notations we denote the $J^{PC}=1^{-+}, 0^{++}, 1^{+-}, 0^{--}$ 
hybrid mesons with up or down quark by $H_1, H_2, H_3, H_4$ 
and those with strange quark by $H^s_1, H^s_2, H^s_3, H^s_4$ respectively.
We define the overlapping amplitude $f_i$ as
\begin{equation}\label{overlap-1}
\langle 0 | J_\mu (0) | H_1\rangle =f_1 \epsilon^1_\mu \; ,
\end{equation}
\begin{equation}\label{overlap-2}
\langle 0 | J_\mu (0) | H_2\rangle =f_2 v_\mu \; ,
\end{equation}
\begin{equation}\label{overlap-3}
\langle 0 | J^5_\mu (0) | H_3\rangle =f_3 \epsilon^3_\mu \; ,
\end{equation}
\begin{equation}\label{overlap-4}
\langle 0 | J^5_\mu (0) | H_4\rangle =i f_4 v_\mu \; ,
\end{equation}
where $\epsilon^1_\mu, \epsilon^3_\mu$ is the $H_1, H_3$ polarization vectors. 

The dispersion relation for $\Pi_i (\omega )$ reads
\begin{equation}\label{dip-1}
\Pi_i ( \omega )=\int {\rho_i (s)\over s- \omega  -i\epsilon }ds\;,
\end{equation}
where $\rho (s)$ is the spectral density in the limit $m_Q \to \infty$.

At the phenomenological side
\begin{equation}
\Pi_i (\omega )={1\over 2} {f^2_i
\over  \Lambda_i-\omega}+\mbox{excited states}+\mbox{continuum} \;.
\end{equation}

In order to suppress the continuum and higher excited states 
contribution we make Borel transformation with the variable $\omega$ to 
(\ref{dip-1}). We have
\begin{equation}
\label{mass-1}
{1\over 2}f_i^2e^{-{\Lambda_i\over T}} 
=\int_0^{s_0} \rho_i (s) e^{-{s\over T}}ds\;,
\end{equation}
where $s_0$ is the continuum threshold. Starting from $s_0$ we have modeled
the phenomenological spectral density with the parton-like ones including 
both the perturbative term and various condensates.

We need the spectral density $\rho_i (s)$ at the quark level. 
The relevant feynman diagrams for the derivation of $\rho_i (s)$
are depicted in FIG. 1. The first line is the heavy quark propagator.
The broken solid line, broken curly line
and a broken solid line with a curly line attached in the middle
stands for the quark condensate, gluon condensate 
and quark gluon mixed condensate respectively. We consider condensates
with dimension less than seven. The last four
diagrams in FIG. 1 involve with quark gluon mixed condensates. They appear
as $\alpha_s \langle {\bar q}g_s \sigma\cdot G q\rangle$. Compared with the
quark condensate, they are typically suppressed by a factor 
${m_0^2\over 16T^2}$, where $m_0^2={\langle {\bar q}g_s \sigma\cdot G q\rangle
\over \langle {\bar q}q\rangle } =0.8$GeV$^2$ and $T\sim 1$GeV. Its 
contribution is negligible. At dimension six there are two condensates,
the four quark condensate and triple gluon condensate, corresponding to 
the sixth and fourth diagram in FIG. 1. But in the present case the 
four quark condensate appears as $\alpha_s^2 \langle {\bar q} q\rangle^2$.
It is of high order in $\alpha_s$ so it can be omitted safely. 
Recall in the QCD sum rule analysis of the light hybrid meson masses, 
the four quark condensate plays a cruicial role \cite{latorre}. Up 
to now there exists no reliable way to estimate its value. 
Two approaches are commonly used to deal with this problem.
One is to invoke the vacuum saturation hypothesis \cite{svz} and use 
the factorization approximation. The other is to scale the value derived 
from the vacuum saturation hypothesis by a number. In \cite{latorre}
a factor two is used to estimate the four quark condensate value, 
which introduces large uncertainty. In contrast, in the framework of HQET, 
the dominant nonperturbative corrections in the QSR analysis of 
heavy hybrid meson masses are due to the quark condensate and 
gluon condensate, which have been determined rather precisely. 

In the calculation we need the following formulas for the gluon condensates.
\begin{equation}
<g_s^2 G_{\alpha\beta}^m G_{\mu\nu}^n >={\delta^{mn}\over 96}
(g_{\alpha\mu}g_{\beta\nu}-g_{\alpha\nu}g_{\beta\mu})
<g_s^2 G^2>\;,
\end{equation}
\begin{eqnarray}\nonumber
<g_s^3f^{abc} G^a_{\mu\nu}G^b_{\alpha\beta}G^c_{\rho\sigma}>
={1\over 24} <g_s^3f^{abc} G^a_{\gamma\delta}G^b_{\delta\epsilon}
G^c_{\epsilon\gamma}>(g_{\mu\sigma}g_{\alpha\nu}g_{\beta\rho} 
+g_{\mu\beta}g_{\alpha\rho}g_{\sigma\nu}&\\
+g_{\alpha\sigma}g_{\mu\rho}g_{\nu\beta}
+g_{\rho\nu}g_{\mu\alpha}g_{\beta\sigma}
-g_{\mu\beta}g_{\alpha\sigma}g_{\rho\nu}
-g_{\mu\sigma}g_{\alpha\rho}g_{\nu\beta}
-g_{\alpha\nu}g_{\mu\rho}g_{\beta\sigma}
-g_{\beta\rho}g_{\mu\alpha}g_{\nu\sigma} )& \;.
\end{eqnarray}
\begin{equation}
Tr\left( {\lambda^a\over 2}{\lambda^b\over 2}{\lambda^c\over 2}\right)
={1\over 4}(d_{abc}+if_{abc}) \;,
\end{equation}
where $d_{abc}, f_{abc}$ are the symmetric and anti-symmetric $SU(3)$
color group structure constants.

The heavy quark propogator has a simple form in coordinate space.
\begin{equation}\label{prop}
<0|T\{h_v (x), {\bar h}_v (0)\}|0>=\int_0^\infty dt \delta (x-vt)\;.
\end{equation}
It's convenient to calculate the Feynman diagrams directly in coordinate space.
Then we perform Wick rotation and make Borel transformation 
with the variable $\omega$ using the Borel transformation formula
\begin{equation}
{\hat {\cal B}}^T_{\omega} e^{\alpha \omega}=\delta (\alpha -{1\over T})\;.
\end{equation}
As a last step we make a second Borel transformation to $\Pi_i (T)$ 
with respect to $\tau ={1\over T}$ to get the spectral density, 
\begin{equation}
\rho_i (s) ={\hat {\cal B}}^s_{\tau} \Pi_i (T=1/\tau ) \; .
\end{equation}

Finally we get
\begin{eqnarray}\label{spectral}\nonumber
\rho (s) ={\alpha_s\over 15\pi^3}s^6+c_1 {\alpha_s\over 15\pi^3}m_q s^5
+c_2 {\alpha_s\over 9\pi^3}a_q s^3
+c_3{1\over 48\pi^2} \langle g^2_sG^2\rangle s^2 &\\
+c_4{1\over 192\pi^2}\langle g^3_sG^3\rangle
+c_5{1\over 192\pi^2}\langle g^2_sG^2\rangle a_q \delta (s) \;, &
\end{eqnarray}
where $a_q=-4\pi^2 \langle {\bar q} q \rangle$, $q=u, d, s$ and 
$\langle g^3_sG^3\rangle =\langle g^3_sf_{abc}
G^a_{\mu\nu}G^b_{\nu\alpha}G^c_{\alpha\mu}\rangle $. 
The coefficients $c_i$ are collected in TABLE I for $H_i$.

The contribution from the nonzero light quark mass is calculated up
to the order ${\cal O} (m_q)$. This term is negligible for sum rules involved
with up and down quark but it is important when the heavy hybrid meson
contains a strange quark as we shall see later.

For the leading order binding energy of heavy hybrid meson we have  
\begin{equation}\label{derivative}
\Lambda_i ={\int_0^{s_0} s \rho (s) e^{-{s\over T}}ds \over
\int_0^{s_0} \rho (s) e^{-{s\over T}}ds } \; .
\end{equation}

\subsection{Numerical analysis of the mass sum rules}

We use $\alpha_s ={4\pi\over (11-{2\over 3}n_f)
\ln ({s_0/2\over \Lambda_{\small \mbox{QCD}}})^2}$ 
and kept four active flavors with $\Lambda_{\mbox{QCD}}=220$ MeV.
The tiny up and down quark mass is taken to be zero.
The quark and gluon condensates adpots the standard values
\begin{eqnarray}
\label{parameter}
\langle\bar uu\rangle&=&-(0.225 ~\mbox{GeV})^3\;,\nonumber\\
\langle\alpha_s G^2\rangle&=&0.038 ~\mbox{GeV}^4\;.
\end{eqnarray}
The value of the triple gluon condensate is not well known. 
In fact several values exist in literature. 
We use \cite{svz}
\begin{equation}\label{ggg}
\langle g^3 G^3\rangle =(1.2 ~\mbox{GeV}^2)\langle\alpha_s G^2\rangle 
=0.045\mbox{GeV}^6\;,
\end{equation}
which is smaller than the value $0.06-0.1$ GeV$^6$ in \cite{zhit} and 
a even larger value $0.4$ GeV$^6$ from the "instanton liquid" approach
to the QCD vacuum \cite{balitsky}.
Later we will enlarge (\ref{ggg}) by a factor of ten to see the uncertainty
from this source.

There are two commonly used methods to extract the masses, 
the derivative method and the fitting method. With the derivative method
we arrive at (\ref{derivative}). The fitting method involves with 
fitting the left hand side (L.H.S.)
and right hand side (R.H.S.) of Eq. (\ref{mass-1}) with the most suitable 
parameters $\Lambda_i, f_i, s_i^0$ directly in the working region
of the Borel parameter. In the numerical analysis, we invoke both 
methods to crosscheck our results. We find both methods 
yield nearly the same results.

We require that (1) the absolute value of each condensate contribution
be less than $30\%$ of the leading perturbative term with continuum 
subtracted and (2) the sum of the power corrections be less than 
one third of the whole sum rule. This requirement leads to the 
lower limit of the continuum threshold. Typically the triple gluon 
condensate is less than $0.5\%$ of the leading term with the 
value in (\ref{ggg}). The series of the operator expansion converge fast. 

We present the numerical values of $\Lambda, f, s_0$ in TABLE II. 
It is understood that there is an error of $0.1$ GeV for $\Lambda, s_0$.
With these values the left hand side and right hand side of 
(\ref{mass-1}) agree within one percent in a large interval of $T$.
Correspondingly, the ratio between the quark, gluon, 
triple gluon condensate and the perturbative term is
collected in TABLE III. 

The dependence of the binding energy on the Borel parameter $T$ 
with different continuum threshold $s_0$ is shown in FIG. 2 and 3
for the two exotic hybrid mesons.
We have also plotted the fitting lines and the curves of the right hand 
side of Eq. (\ref{mass-1}) for $H_1, H_4$ in FIG. 4 and 5. 

Varying $\Lambda_{\mbox{QCD}}$ from 
$220$MeV to $300$MeV the final result changes within $5\%$.
in our numerical analysis. 

If we use $\langle g^3 G^3\rangle =0.45\mbox{GeV}^6$, 
the values of $\Lambda, s_0$ is shifted upwards by $2\%$ for $H_1$. 
For the other three channels the best fitting parameters changes little. 
In other words, the uncertainty due to the triple gluon condensate 
is small and included in the errors given already. This is in 
strong contrast with the case for the light hybrid mesons, where 
the hybrid masses depend cruicially on the value of the triple gluon
condensate. Varying its value from $0.045\mbox{GeV}^6$ to $0.4\mbox{GeV}^6$,
the $1^{-+}$ hybrid meson mass increases from $1.0$ GeV to $1.5$ GeV
\cite{balitsky}.

For the numerical analysis of the strange heavy hybrid meson, 
we use $\langle\bar ss\rangle =0.8 \langle\bar uu\rangle$, 
$m_s =150$ MeV. The results are collected in TABLE IV and V.
As can be seen from TABLE V the strange quark mass correction 
is very important. The binding 
energy of the light component of $H_i^s$ is roughly $130$ MeV larger than 
that for $H_i$. The continuum threshold increases by about $150$ MeV. 
The decay constants increase typically by $25\%$ due to the strange quark 
mass correction.

It is interesting to note that the central 
value of $\Lambda_H$ is much greater than that for 
the $(0^-, 1^-)$ doublet of $Q\bar q$ meson, 
$\Lambda_{-,{1\over 2}}=(0.5\pm 0.1) $GeV.
And the continuum starts at rather large $s_0$
due to the presence of the dynamical gluon in $H_Q$.  

The masses of heavy hybrid mesons shall be around $(m_b +\Lambda_H)$. 
If we can derive $\Lambda_H$ reliably, we have a good estimate of $H_Q$ mass.
Especially for the bottom quark system, the $1/m_Q$ correction is not large.
This point has been the motivation of our considering heavy  
hybrid mesons with one heavy quark in the framework of HQET. 

The uncertainty due to the 
dimesion six condensates renders the reliable extraction of 
the light hybrid meson masses rather difficult. 
In this section we have calculated the binding energy of the heavy  
hybrid meson masses in the leading order of HQET with QCD sum rules. 
Within the present approach (1) the heavy quark mass is disentangled;
(2) the four quark condensate is of higher order in $\alpha_s$, hence 
can be neglected safely; (3) the dominant power corrections are 
from the quark condensate and the gluon condensate, which is 
well known. The triple gluon condensate shall at most affect the 
binding energy by $3\%$ varying its value 
from $0.045\mbox{GeV}^6$ to $0.4\mbox{GeV}^6$.

\section{Light cone QCD sum rules for the pionic couplings}
\label{sec3}
\setcounter{equation}{0}
Note $H_1$ is the lightest exotic heavy hybrid meson while $H_4$
lies about $1.4$ GeV higher than $H_1$. Experimental discovery 
of the $0^{++}$ and $1^{+-}$ heavy hybrid mesons will be difficult
since they have the same non-exotic quantum numbers as the radial 
excitations of ordinary heavy mesons. In this section we discuss
the decay modes and widths of the loweset heavy hybrid meson $H_1$.

Denote the doublet $(1^+,2^+)$ with $j_{\ell}=3/2$ by $(B_1,B_2^*)$, 
the doublet $(0^+,1^+)$ with $j_{\ell}=1/2$ by $(B^{\prime}_0,B^{\prime}_1)$ 
and the doublet $(0^-, 1^-)$ by $(B, B^*)$. 

We need the following interpolating currents \cite{huang}
\begin{eqnarray}
\label{curr1}
&&J^{\dag\alpha}_{1,+,{3\over 2}}=\sqrt{\frac{3}{2}}\:\bar h_v\gamma^5(-i)\left(
{\cal D}_t^{\alpha}-\frac{1}{3}\gamma_t^{\alpha} {\cal D}_t\right)q\;,\\
\label{curr2}
&&J^{\dag\alpha_1,\alpha_2}_{2,+,{3\over 2}}=\:\bar h_v
\frac{(-i)}{2}\left(\gamma_t^{\alpha_1}{\cal D}_t^{\alpha_2}+
\gamma_t^{\alpha_2}{\cal D}_t^{\alpha_1}-{2\over 3}g_t^{\alpha_1\alpha_2}
 {\cal D}_t\right)q\;,\\
\label{curr3}
&&J^{\dag\alpha}_{1,-,{1\over 2}}=\:\bar h_v\gamma_t^{\alpha}
q\;,\hspace{1.5cm} J^{\dag\alpha}_{0,-,{1\over 2}}=\:\bar h_v\gamma_5q\;,
\end{eqnarray}
\begin{eqnarray}
\label{current1}
J^{\dag}_{0,+,{1\over 2}}=\:\bar h_vq\;,\hspace{1.5cm}
J^{\dag\alpha}_{1,+,{1\over 2}}=\:\bar h_v\gamma^5\gamma^{\alpha}_tq\;,
\end{eqnarray} 
where $\gamma_t^\mu =\gamma^\mu -{\hat v}v^\mu$, 
$g^t_{\mu\nu}=g_{\mu\nu}-v_\mu v_\nu$. We use "t" to denote the 
tansverse index. 

We take the p-wave decay process $H_1 \to B \pi $ as an example to
illustrate the method.
\begin{itemize}
\item $H_1 \to B \pi $

The decay amplitude is 
\begin{equation}\label{amp-1}
 M(H_1\to B \pi  )= \epsilon_1^\mu q^t_\mu g_1 \;,
\end{equation}
with $q^t_\mu =q_\mu -(q\cdot v) v_\mu$.

For deriving the sum rules for the coupling constant $g_1$
we consider the correlator 
\begin{equation}\label{cor-1}
i \int d^4x\;e^{ik\cdot x}\langle\pi (q)|T\left(J_{0,-,\frac{1}{2}}(x)
 J^{\dagger\mu}(0)\right)|0\rangle 
 = G_1 (\omega,\omega')q^t_\mu\;. 
\end{equation}

The function $G_1 (\omega,\omega^{\prime})$ in (\ref{cor-1})
has the following double dispersion relation 
\begin{eqnarray}
\label{pole-1}
{f_{-,{1\over 2}}f_H g_1\over 4(\Lambda_{-,{1\over 2}}
-\omega)(\Lambda_H -\omega')}+{c\over \Lambda_{-,{1\over 2}}
-\omega}+{c'\over \Lambda_H-\omega'}\;,
\end{eqnarray}
where $\Lambda_{P,j_\ell}=m_{P,j_\ell}-m_Q$ and 
$f_{P,j_\ell}$ are constants defined as:
\begin{equation}
\langle 0|J_{j,P,j_{\ell}}^{\alpha_1\cdots\alpha_j}(0)|j',P',j_{\ell}^{'}\rangle=
f_{Pj_l}\delta_{jj'}
\delta_{PP'}\delta_{j_{\ell}j_{\ell}^{'}}\eta^{\alpha_1\cdots\alpha_j}\;.
\end{equation}

Keeping the three particle component of the pion wave function, the expression
for $G_1 (\omega, \omega')$ with tensor structure reads
\begin{equation}\label{g-1}
G_1 (\omega, \omega')= \int d^4x e^{ikx} \delta (x-vt)
{\bf Tr}\{ \gamma_5 {1+{\hat v}\over 2} \gamma^\nu <\pi (q) |
q(0) g_s G_{\mu\nu}(0) {\bar q}(x) |0> \} \; ,
\end{equation}
where we have not included the two particle component of the pion wave function 
since they are of higher order in $\alpha_s$ and suppressed by a large 
factor ${g_s^2\over (4\pi)^2}$, which arises from the additional loop 
integration with the gluon attached to one of the quark line.

The light cone three particle pion wave functions are defined as \cite{bely-95}:
\begin{eqnarray}
& &<\pi(q)| {\bar d} (x) \sigma_{\alpha \beta} \gamma_5 g_s 
G_{\mu \nu}(ux)u(0) |0>=
\nonumber \\ &&i f_{3 \pi}[(q_\mu q_\alpha g_{\nu \beta}-q_\nu q_\alpha g_{\mu \beta})
-(q_\mu q_\beta g_{\nu \alpha}-q_\nu q_\beta g_{\mu \alpha})]
\int {\cal D}\alpha_i \; 
\varphi_{3 \pi} (\alpha_i) e^{iqx(\alpha_1+v \alpha_3)} \;\;\; ,
\label{p3pi} 
\end{eqnarray}

\begin{eqnarray}
& &<\pi(q)| {\bar d} (x) \gamma_{\mu} \gamma_5 g_s 
G_{\alpha \beta}(vx)u(0) |0>=
\nonumber \\
&&f_{\pi} \Big[ q_{\beta} \Big( g_{\alpha \mu}-{x_{\alpha}q_{\mu} \over q \cdot 
x} \Big) -q_{\alpha} \Big( g_{\beta \mu}-{x_{\beta}q_{\mu} \over q \cdot x} 
\Big) \Big] \int {\cal{D}} \alpha_i \varphi_{\bot}(\alpha_i) 
e^{iqx(\alpha_1 +v \alpha_3)}\nonumber \\
&&+f_{\pi} {q_{\mu} \over q \cdot x } (q_{\alpha} x_{\beta}-q_{\beta} 
x_{\alpha}) \int {\cal{D}} \alpha_i \varphi_{\|} (\alpha_i) 
e^{iqx(\alpha_1 +v \alpha_3)} \hskip 3 pt  \label{gi} 
\end{eqnarray}
\noindent and
\begin{eqnarray}
& &<\pi(q)| {\bar d} (x) \gamma_{\mu}  g_s \tilde G_{\alpha \beta}(vx)u(0) |0>=
\nonumber \\
&&i f_{\pi} 
\Big[ q_{\beta} \Big( g_{\alpha \mu}-{x_{\alpha}q_{\mu} \over q \cdot 
x} \Big) -q_{\alpha} \Big( g_{\beta \mu}-{x_{\beta}q_{\mu} \over q \cdot x} 
\Big) \Big] \int {\cal{D}} \alpha_i \tilde \varphi_{\bot}(\alpha_i) 
e^{iqx(\alpha_1 +v \alpha_3)}\nonumber \\
&&+i f_{\pi} {q_{\mu} \over q \cdot x } (q_{\alpha} x_{\beta}-q_{\beta} 
x_{\alpha}) \int {\cal{D}} \alpha_i \tilde \varphi_{\|} (\alpha_i) 
e^{iqx(\alpha_1 +v \alpha_3)} \hskip 3 pt . \label{git} 
\end{eqnarray}
\noindent 
The operator $\tilde G_{\alpha \beta}$  is the dual of $G_{\alpha \beta}$:
$\tilde G_{\alpha \beta}= {1\over 2} \epsilon_{\alpha \beta \delta \rho} 
G^{\delta \rho} $; ${\cal{D}} \alpha_i$ is defined as 
${\cal{D}} \alpha_i =d \alpha_1 
d \alpha_2 d \alpha_3 \delta(1-\alpha_1 -\alpha_2 
-\alpha_3)$. 
Due to the choice of the
gauge  $x^\mu A_\mu(x) =0$, the path-ordered gauge factor
$P \exp\big(i g_s \int_0^1 du x^\mu A_\mu(u x) \big)$ has been omitted.

The function $\varphi_{3 \pi}$ is of twist three, while all the wave 
functions appearing in eqs.(\ref{gi}), (\ref{git}) are of twist four.
The wave functions $\varphi (x_i,\mu)$ ($\mu$ is the renormalization point) 
describe the distribution in longitudinal momenta inside the pion, the 
parameters $x_i$ ($\sum_i x_i=1$) 
representing the fractions of the longitudinal momentum carried 
by the quark, the antiquark and gluon.

The wave function normalizations immediately follow from the definitions
$\int {\cal D} \alpha_i \varphi_{3\pi}(\alpha_i)=1$,
$\int {\cal D} \alpha_i \varphi_\bot(\alpha_i)=
\int {\cal D} \alpha_i \varphi_{\|}(\alpha_i)=0$,
$\int {\cal D} \alpha_i \tilde \varphi_\bot(\alpha_i)=-
\int {\cal D} \alpha_i \tilde \varphi_{\|}(\alpha_i)={\delta^2/3}$,
with the parameter $\delta$ defined by 
the matrix element: 
$<\pi(q)| {\bar d} g_s \tilde G_{\alpha \mu} \gamma^\alpha u |0>=
i \delta^2 f_\pi q_\mu$.

Expressing (\ref{g-1}) with the pion wave functions we arrive at:
\begin{equation}\label{g-11}
G_1 (\omega, \omega')= -\int_0^{\infty} dt\int {\cal D} \alpha_i
e^{i(kv+qv\alpha_1)t} \{f_{3\pi} \varphi_{3\pi}(\alpha_i) (q\cdot v)
+f_\pi [\varphi_\bot(\alpha_i)-{\varphi_{\|}(\alpha_i)\over 2}] \} +\cdots\; .
\end{equation}

For large Euclidean values of $\omega$ and $\omega'$ 
this integral is dominated by the region of small $t$, therefore it can be 
approximated by the first a few terms.

After Wick rotations and making double Borel transformation 
with the variables $\omega$ and $\omega'$
the single-pole terms in (\ref{pole-1}) are eliminated. 
We arrive at:
\begin{equation}\label{g1}
{1\over 4} g_1 f_{-,{1\over 2} } f_H  
e^{-( { \Lambda_{-,{1\over 2} } \over T_1}+{ \Lambda_H \over T_2} )}= 
 f_{3\pi} \Phi'_{3\pi}(u_0) 
T^2 -f_\pi [\Phi_\bot(u_0)-{\Phi_{\|}(u_0)\over 2}] T \;,
\end{equation}
where $f_H=f_1$, $u_0={T_1 \over T_1 + T_2}$, 
$T\equiv {T_1T_2\over T_1+T_2}$, $T_1$, $T_2$ are the Borel parameters.
Note the sum rule is asymmetric
with the Borel parameter $T_1$ and $T_2$. The continuum subtraction
is complicated in the present case. We shall discuss this point in 
section \ref{continuum}. The right hand side of the sum rule 
is the result after integration of the double spectral density 
with respect to $s_1$, $s_2$ in the interval $(0, \infty)$ so 
it includes the continuum contribution.

The new wave functions introduced in (\ref{g1}) are defined as
\begin{equation}
\Phi_i (\alpha_1 )=\int_0^{1-\alpha_1} \varphi_i (\alpha_1, \alpha_2,
1-\alpha_1-\alpha_2 ) d\alpha_2 \;,
\end{equation}
with $i=3\pi, \bot, {\|}$ etc and
\begin{equation}
\Phi'_{3\pi} (u_0 )={d\Phi_{3\pi} (\alpha_1 )\over d\alpha_1}|_{\alpha_1=u_0} \;.
\end{equation}

We have used the Borel transformation formula:
${\hat {\cal B}}^T_{\omega} e^{\alpha \omega}=\delta (\alpha -{1\over T})$.
Integration by parts are employed to absorb the factor $(q\cdot v)$, 
which leads to the derivative in (\ref{g1}). In this way we 
arrive at the simple form after double Borel transformation.

\item $H_1 \to B^* \pi $

\begin{equation} \label{1-2}
 M(H_1\to B^* \pi  )= i\epsilon_{\mu\alpha\sigma\beta}
\epsilon^\mu_1 \epsilon^\alpha q^\sigma v^\beta g_2  \;,
\end{equation}
where $\epsilon_{\mu}$ is the polarization vector of $B^*$.

Similarly we consider the correlator 
\begin{equation}\label{cor-2}
i \int d^4x\;e^{ik\cdot x}\langle\pi (q)|T\left(J^\alpha_{1,-,\frac{1}{2}}(x)
 J^{\dagger\mu}(0)\right)|0\rangle 
= i\epsilon_{\mu\alpha\sigma\beta} q^\sigma v^\beta G_2 (\omega,\omega') \;. 
\end{equation}

\begin{equation}\label{g-22}
G_2 (\omega, \omega')= {f_\pi\over 2}
 \int_0^{\infty} dt\int {\cal D} \alpha_i e^{i(kv+qv\alpha_1)t} 
[\varphi_\bot(\alpha_i) -{\tilde \varphi}_\bot(\alpha_i)
+{\tilde \varphi}_{\|}(\alpha_i)]  +\cdots\; .
\end{equation}

\begin{equation}\label{g2}
{1\over 4} g_2 f_{-,{1\over 2} } f_H  
e^{-( { \Lambda_{-,{1\over 2} } \over T_1}+{ \Lambda_H \over T_2} )}= 
-{f_\pi\over 2} [\Phi_\bot (u_0)-{\tilde \Phi}_\bot (u_0)
+{\tilde\Phi}_{\|}(u_0)] T  \;.
\end{equation}

\item $H_1 \to B'_0 \pi$

This process is forbidden due to the parity and angular momentum conservation.

\item $H_1 \to B'_1 \pi$

There exist two independent coupling constants, corresponding to S-wave and 
D-wave decay. The decay amplitudes are:
\begin{equation} \label{1-3}
 M(H_1\to B'_1 \pi  )= \eta^\alpha\epsilon_1^\mu \{ g^t_{\mu\alpha} g_3
 +(q^t_\alpha q^t_\mu -q^2_t g^t_{\mu\alpha}) g_4 \}   \;.
\end{equation}
where $\eta_{\mu}$ is the polarization vector of $B^*$.

We consider the correlator 
\begin{equation}\label{cor-3}
i \int d^4x\;e^{ik\cdot x}\langle\pi (q)|T\left(J^\alpha_{1,+,\frac{1}{2}}(x)
 J^{\dagger\mu}(0)\right)|0\rangle 
=  g^t_{\mu\alpha} G_3 (\omega,\omega')
+(q^t_\alpha q^t_\mu -q^2_t g^t_{\mu\alpha})G_4 (\omega,\omega') \;. 
\end{equation}

\begin{equation}\label{g-33}
G_3 (\omega, \omega')= \int_0^{\infty} dt\int {\cal D} \alpha_i
e^{i(kv+qv\alpha_1)t}(q\cdot v) \{-f_{3\pi} \varphi_{3\pi}(\alpha_i) (q\cdot v)
+f_\pi [{\tilde\varphi}_\bot(\alpha_i)+{\varphi_{\|}(\alpha_i)\over 2}] \}
+\cdots \; .
\end{equation}

\begin{eqnarray}\label{g-44}\nonumber
G_4 (\omega, \omega')= \int_0^{\infty} dt\int {\cal D} \alpha_i
e^{i(kv+qv\alpha_1)t}\{f_{3\pi} \varphi_{3\pi}(\alpha_i) (q\cdot v)
+{1\over 2}{f_\pi\over q\cdot v} 
[\varphi_\bot (\alpha_i)&\\
+\varphi_{\|}(\alpha_i)
+{\tilde\varphi}_\bot (\alpha_i)+{\tilde\varphi}_{\|}(\alpha_i)] \}
+\cdots \; . &
\end{eqnarray}

\begin{equation}\label{g3}
{1\over 4} g_3 f_{+,{1\over 2} } f_H  
e^{-( { \Lambda_{+,{1\over 2} } \over T_1}+{ \Lambda_H \over T_2} )}= 
- f_{3\pi} {d^2\Phi_{3\pi}(\alpha_1) \over d\alpha_1^2}|_{\alpha_1 =u_0}
T^3 
-f_\pi [\Phi'_{\|} (u_0)+{{\tilde\Phi}'_\bot (u_0)\over 2}]T^2  \;.
\end{equation}

\begin{equation}\label{g4}
{1\over 4} g_4 f_{+,{1\over 2} } f_H  
e^{-( { \Lambda_{+,{1\over 2} } \over T_1}+{ \Lambda_H \over T_2} )}= 
 f_{3\pi} \Phi_{3\pi}(u_0) T +{f_\pi\over 2} \Phi (u_0) \;,
\end{equation}
where 
\begin{equation}
\Phi (u) =\int_0^u d\alpha [\varphi_\bot (\alpha)+\varphi_{\|}(\alpha)
+{\tilde\varphi}_\bot (\alpha)+{\tilde\varphi}_{\|}(\alpha)]\;.
\end{equation}

\item $H_1 \to B_1 \pi$

There also exist two independent coupling constants, corresponding to S-wave and 
D-wave decay. The decay amplitudes are:
\begin{equation} \label{1-5}
 M(H_1\to B_1 \pi  )= \eta^\alpha\epsilon_1^\mu \{ g^t_{\mu\alpha} g_5
 +(q^t_\alpha q^t_\mu -q^2_t g^t_{\mu\alpha}) g_6 \}   \;.
\end{equation}
where $e_{\mu}$ is the polarization vector of $B_1$.

We consider the correlator 
\begin{equation}\label{cor-5}
i \int d^4x\;e^{ik\cdot x}\langle\pi (q)|T\left(J^\alpha_{1,+,\frac{3}{2}}(x)
 J^{\dagger\mu}(0)\right)|0\rangle 
=  g^t_{\mu\alpha} G_5 (\omega,\omega')
+(q^t_\alpha q^t_\mu -q^2_t g^t_{\mu\alpha})G_6 (\omega,\omega') \;. 
\end{equation}

\begin{eqnarray}\label{g-55}\nonumber
G_5 (\omega, \omega')=
-{\sqrt{6}\over 3} \int_0^{\infty} dt\int {\cal D} \alpha_i
e^{i(kv+qv\alpha_1)t}\alpha_1 (q\cdot v)^2 
\{f_{3\pi} \varphi_{3\pi}(\alpha_i) (q\cdot v) &\\
+f_\pi [\varphi_\bot (\alpha_i)-{\varphi_{\|}(\alpha_i)\over 2}] \}
+\cdots \; .&
\end{eqnarray}

\begin{eqnarray}\label{g-66}\nonumber
G_6 (\omega, \omega')= 
{\sqrt{6}\over 12}\int_0^{\infty} dt\int {\cal D} \alpha_i
e^{i(kv+qv\alpha_1)t}\alpha_1 \{4 f_{3\pi} \varphi_{3\pi}(\alpha_i) (q\cdot v)
+f_\pi [5\varphi_\bot (\alpha_i)&\\
-2\varphi_{\|}(\alpha_i)
-{\tilde\varphi}_\bot (\alpha_i)+{\tilde\varphi}_{\|}(\alpha_i)] \}
+\cdots \; .&
\end{eqnarray}

\begin{eqnarray}\label{g5}\nonumber
{1\over 4} g_5 f_{+,{3\over 2} } f_H  
e^{-( { \Lambda_{+,{3\over 2} } \over T_1}+{ \Lambda_H \over T_2} )}= 
{\sqrt{6}\over 3}
\{ f_{3\pi} 
{d^3 [\alpha_1\Phi_{3\pi}(\alpha_1) ]\over d\alpha_1^3}|_{\alpha_1 =u_0}
T^4 &\\
-f_\pi {d^2[\alpha_1\Phi_\bot (\alpha_1) -{\alpha_1\Phi_{\|} (\alpha_1)\over 2}]
\over d\alpha_1^2}|_{\alpha_1 =u_0} T^3 \} \;.&
\end{eqnarray}

\begin{eqnarray}\label{g6}\nonumber
{1\over 4} g_6 f_{+,{3\over 2} } f_H  
e^{-( { \Lambda_{+,{3\over 2} } \over T_1}+{ \Lambda_H \over T_2} )}= 
-{\sqrt{6}\over 12}
\{ 4f_{3\pi}{d[\alpha_1 \Phi_{3\pi}(\alpha_1)]\over d\alpha_1}|_{\alpha_1 =u_0}
T^2 &\\ -f_\pi u_0[ 5\Phi_\bot (u_0) -2\Phi_{\|}(u_0)
-{\tilde\Phi}_\bot (u_0)+{\tilde\Phi}_{\|}(u_0)] T \} \;.&
\end{eqnarray}

\item $H_1\to B_2^* \pi$

There exists only one independent coupling constant, corresponding to  
D-wave decay. The decay amplitude is:
\begin{equation} \label{1-7}
 M(H_1\to B_2^* \pi  )= \eta^{\alpha_1\alpha_2}\epsilon^1_\mu 
[i\epsilon^{\alpha_1\sigma\beta\mu}v_\sigma q_\beta q^t_{\alpha_2} 
+(\alpha_1 \leftrightarrow \alpha_2 ) ] g_7 \;.
\end{equation}
where $\eta_{\alpha_1\alpha_2}$ is the polarization tensor of $B_2^*$.

We consider the correlator 
\begin{equation}\label{cor-7}
i \int d^4x\;e^{ik\cdot x}\langle\pi (q)|T\left(J^\alpha_{2,+,\frac{3}{2}}(x)
 J^{\dagger\mu}(0)\right)|0\rangle 
= [i\epsilon^{\alpha_1\sigma\beta\mu}v_\sigma q_\beta q^t_{\alpha_2} 
+(\alpha_1 \leftrightarrow \alpha_2 ) ] G_7 (\omega,\omega') \;. 
\end{equation}

\begin{equation}\label{g-77}
G_7 (\omega, \omega')= 
{f_\pi\over 4}\int_0^{\infty} dt\int {\cal D} \alpha_i
e^{i(kv+qv\alpha_1)t}\alpha_1 [\varphi_\bot (\alpha_i)
-{\tilde\varphi}_\bot (\alpha_i)+{\tilde\varphi}_{\|}(\alpha_i)]
+\cdots \; .
\end{equation}

\begin{equation}\label{g7}
{1\over 4} g_7 f_{+,{3\over 2} } f_H  
e^{-( { \Lambda_{+,{3\over 2} } \over T_1}+{ \Lambda_H \over T_2} )}= 
{f_\pi\over 4} u_0 [\Phi_\bot (u_0)
-{\tilde\Phi}_\bot (u_0)+{\tilde\Phi}_{\|}(u_0)]T \;.
\end{equation}

\end{itemize}

\section{Determination of the parameters}
\label{sec4} 
\setcounter{equation}{0}
\subsection{The values of $\Lambda, f$}

We need the mass parameters $\Lambda$'s and the coupling constants 
$f$'s of the corresponding interpolating currents as input. 
The results are \cite{neubert,zhu-duality}
\begin{equation}\label{lambda-1}
\Lambda_{-,{1\over 2}}=(0.5\pm 0.10)\mbox{GeV}\;,
\end{equation}
\begin{equation}\label{f-1}
f_{-,{1\over 2}}=(0.35\pm 0.04)\mbox{GeV}^{3\over 2}\;,
\end{equation}
\begin{equation}\label{lambda-2}
\Lambda_{+,{1\over 2}}=(0.85\pm 0.10)\mbox{GeV}\;,
\end{equation}
\begin{equation}\label{f-2}
f_{+,{1\over 2}}=(0.36\pm 0.04)\mbox{GeV}^{3\over 2}\;,
\end{equation}
\begin{equation}\label{lambda-3}
{\bar \Lambda}_{+,{3\over 2}}=(0.95\pm 0.10)\mbox{GeV}\;,
\end{equation}
\begin{equation}\label{f-3}
f_{+,{3\over 2}}=(0.28\pm 0.03)\mbox{GeV}^{5\over 2}\;.
\end{equation}

The value of the continuum threshold for the $(0^-, 1^-), (0^+, 1^+), 
(1^+, 2^+)$ doublets is $(1.1\pm 0.1), (1.2\pm 0.1), (1.3\pm 0.1)$ GeV
respectively.

\subsection{Expressions of PWFs}

The detailed expressions of the pion wave functions relevant in our calculation are:
\begin{equation}\label{fun-a}\nonumber
\varphi_{3\pi}(\alpha_i) =360\alpha_1\alpha_2\alpha_3^2
 [1 +{\omega_{1,0} \over 2}
(7\alpha_3-3) ]  \; ,
\end{equation}
\begin{equation}\label{fun-b}
\varphi_\bot (\alpha_i) =30\delta^2 (\alpha_1 -\alpha_2) \alpha_3^2 [{1\over 3}
 +2\epsilon(1-2\alpha_3)] \; ,
 \end{equation}
\begin{equation}\label{fun-c}
\varphi_{\|} (\alpha_i) =120\delta^2 \epsilon (\alpha_1 -\alpha_2) 
\alpha_1\alpha_2\alpha_3 \; ,
 \end{equation}
\begin{equation}\label{fun-d}
\tilde\varphi_{\bot} (\alpha_i) =30\delta^2 (1 -\alpha_3) \alpha_3^2 [{1\over 3}
 +2\epsilon (1-2\alpha_3)] \; ,
 \end{equation}
\begin{equation}\label{fun-e}
\tilde\varphi_{\|} (\alpha_i) =-120\delta^2 \alpha_1 \alpha_2 \alpha_3 [{1\over 3}
 +\epsilon(1-2\alpha_3)] \; ,
\end{equation}
where the coefficients $f_{3\pi}$, $\omega_{1,0}$ \cite{f-omega}, 
$\delta^2$ \cite{delta2} and $\epsilon$ \cite{epsilon} have been 
determined from QCD sum rules.
At the scale $\mu =1.0$ GeV, $\omega_{1,0}=-2.88$, 
$f_{3\pi}=0.0035$GeV$^2$, $\delta^2 =0.2$GeV$^2$, 
$\epsilon =0.5$. 

With (\ref{fun-a})-(\ref{fun-e}) we can calculate the explicit expressions
of the other wave functions defined in this work. Note only the asymptotic
form of these wave functions are exactly known.  The pieces involved 
with $\epsilon$, $\omega_{1,0}$ arise from nonperturbative corrections.
They are estimated from the moments of the pion wave functions with 
QCD sum rules. For our purpose it's enough to keep the asymptotic form. 
\begin{equation}
\Phi_{3\pi} (u)=30u (1-u)^4 \; ,
\end{equation}
\begin{equation}
[\Phi_{3\pi} (u)]'=30 (1-5u) (1-u)^3 \; ,
\end{equation}
\begin{equation}
[u \Phi_{3\pi} (u)]'=60 u(1-3u) (1-u)^3 \; ,
\end{equation}
\begin{equation}
[\Phi_{3\pi} (u)]^{''}=-120 (2-5u) (1-u)^2 \; ,
\end{equation}
\begin{equation}
[u\Phi_{3\pi} (u)]^{'''}=720 (u-1)(5u^2-5u+1) \; ,
\end{equation}
\begin{equation}
\Phi_\bot (u)={5\over 6}\delta^2 (5u-1) (1-u)^3 \; ,
\end{equation}
\begin{equation}
[u\Phi_\bot (u)]^{''}=-{10\over 3}\delta^2 (25u^3-48u^2+27u-4) \; ,
\end{equation}
\begin{equation}
\Phi_{\|} (u)= [\Phi_{\|} (u)]'=[u\Phi_{\|} (u)]^{''}=0\; ,
\end{equation}
\begin{equation}
{\tilde \Phi}_\bot (u)={5\over 6}\delta^2 (3u+1) (1-u)^3 \; ,
\end{equation}
\begin{equation}
[{\tilde\Phi}_\bot (u)]^{'}=-10\delta^2 u(1-u)^2 \; ,
\end{equation}
\begin{equation}
{\tilde\Phi}_{\|} (u)=-{20\over 3}\delta^2 u (1-u)^3 \; ,
\end{equation}
\begin{equation}
[{\tilde\Phi}_{\|} (u)]'=-{20\over 3}\delta^2 (1-4u) (1-u)^2 \; ,
\end{equation}
\begin{equation}
\Phi (u)=0\;,
\end{equation}
where $', ^{''}, ^{'''}$ denotes the first, second and third derivative 
with respect to $u$.

\section{Subtraction of the continuum}
\label{continuum}
\setcounter{equation}{0}

We have the double dispersion relation for $G_i (\omega, \omega')$,
\begin{eqnarray}\nonumber
\Pi (\omega, \omega')=\int_0^\infty ds_1 \int_0^\infty ds_2 
{\rho (s_1, s_2) \over (s_1-\omega-i\epsilon) (s_2-\omega' -i\epsilon)} 
+\int_0^\infty ds_1 {\rho_1 (s_1) \over (s_1-\omega-i\epsilon) }&\\
+\int_0^\infty ds_2 {\rho_2 (s_2) \over (s_2-\omega-i\epsilon) }
+\cdots \;,&
\end{eqnarray}
where the ellipse denotes the subtraction terms.

Making double Borel transformation to the variables $\omega, \omega'$
we get 
\begin{equation}\label{tau}
\Pi (\tau_1, \tau_2)=\int_0^\infty ds_1 \int_0^\infty ds_2 
e^{-s_1\tau_1}e^{-s_2\tau_2}\rho (s_1, s_2)\;.
\end{equation}
The single pole and subtraction terms have been eliminated in (\ref{tau}).
Making a second double Borel transformation to $\tau_1, \tau_2$,
we get the double spectral density
\begin{equation}
\rho (s_1, s_2)={\cal B}_{\tau_1}^{s_1}{\cal B}_{\tau_2}^{s_2}
\Pi (\tau_1, \tau_2)\;.
\end{equation}

After expressing $G_i (\omega, \omega')$ with the PWFs 
and finishing Wick rotations and double Borel transformation, we 
can get the following general formula:
\begin{eqnarray}\nonumber\label{wsx}
\Pi (\tau_1, \tau_2)=\int_0^\infty dt\int_0^1 
\delta [(1-u)t-\tau_1] \delta (ut-\tau_2) t^{-n} \psi (u) &\\  \nonumber
={1\over (\tau_1 +\tau_2)^{n+1}} \psi ({\tau_2\over \tau-1 +\tau_2})&\\ 
=\sum\limits_{k=0}^{\infty} a_k {{\tau_1}^k\over (\tau_1+\tau_2)^{k+n+1} }  &
\end{eqnarray}
if we assume
\begin{equation}
\psi (u)=\sum\limits_{k=0}^{\infty} a_k (1-u)^k \;.
\end{equation}
In order to faciliate the numerical analysis we collect the coefficients 
$a_k$ in TABLE VI after expanding the PWFs into the polynomials 
of $(1-u)$.

\begin{equation}
\rho(s_1, s_2)=\sum\limits_{k=0}^{\infty} a_k 
{s_2^{k+n}\over \Gamma (k+n+1)} (-{\partial\over \partial s_2})^k 
\delta (s_2 -s_1 ) \;.
\end{equation}

Introducing new variables $s_+={s_1+s_2\over 2}$, $s_+={s_1+s_2\over 2}$,
${1\over T_-}={1\over T_1}-{1\over T_2}$, we have
\begin{equation}\label{strange}
\Pi (T_1, T_2) =2\int_0^\infty ds_+ e^{-{s_+\over T}} 
\int_{-s_+}^{s_+} ds_- e^{s_-\over T_-} \rho (s_+, s_-) \;,
\end{equation}
where 
\begin{equation}
\rho (s_+, s_-)=\sum\limits_{k=0}^{\infty} {a_k \over 2^k}
{(s_++s_-)^{k+n}\over \Gamma (k+n+1)} (-{\partial\over \partial s_-})^k 
\delta (2s_-) \;.
\end{equation}

The general quark-hadron duality holds only after the integration
with respect to the variable $s_-$ in (\ref{strange}). Finally we get
\begin{equation}\label{qaz}
\Pi (T_1, T_2) =\sum\limits_{k=0}^{\infty}\sum\limits_{q=0}^{k}
{a_k \over 2^k} {k!\over (k-q)! q! (k+n-q)!}({1\over T_-})^{k-q}
\int_0^{E_c} s_+^{k+n-q} e^{-{s_+\over T}} ds_+ \;,
\end{equation}
where $E_c$ is the continuum threshold. From its definition we 
know that $E_c ={s_0 +s_Q\over 2}$ where $s_0, s_Q$ is
the continuum threshold for the mass sum rules of the hybrid and heavy
meson respectively.

Letting $E_c\to \infty$ we recover (\ref{wsx}). In the case of symmetric
sum rules, only the term with $q=k$ survives since $T_1=T_2=2T$ and
${1\over T_-}=0$. It is straightforward to get 
\begin{equation}
\Pi (T) =\psi ({1\over 2}) 
\int_0^{E_c} s_+^n e^{-{s_+\over T}} ds_+ \;.
\end{equation}
Now the subtraction of the continuum contribution does not
depend on the wave functions. 

In the present case the sum rules are asymmetric with $T_1$ and $T_2$.
It's reasonable to let $T_1 =2\beta \Lambda_Q$, $T_2=2\beta \Lambda_H$,
where $\Lambda_Q $ is the binding energy of the heavy meson in the 
leading order of HQET and $\beta$ is the dimensionless scale parameter.
Then we have $u_0 ={\Lambda_Q\over \Lambda_H +\Lambda_Q}$, 
$T={2\Lambda_Q\Lambda_H\over \Lambda_Q +\Lambda_H}\beta$. 

We rewrite (\ref{qaz}) as 
\begin{equation}\label{edc}
\Pi (T) =T^{n+1}\sum\limits_{k=0}^{\infty}{a_k \over 2^k} \{
\sum\limits_{q=0}^{k} {k!\over (k-q)! q! }({T\over T_-})^q
f_{n+q} ({E_c\over T}) \} \;,
\end{equation}
where $f_n(x)=1-e^{-x}\sum\limits_{k=0}^{n}{x^k\over k!}$ is the factor 
used to subtract the continuum. Note ${T\over T_-} ={\Lambda_H -\Lambda_Q
\over \Lambda_H +\Lambda_Q}\approx {1\over 4}, {1\over 3}, {1\over 2}$ for the 
$(1^+,2^+)$, $(0^+,1^+)$, $(0^-, 1^-)$ doublet respectively. 
So only the first few terms in the bracket in (\ref{edc}) is important.
Replacing $f_{n+q} ({E_c\over T})$ by $f_n ({E_c\over T})$, we have
\begin{eqnarray}\label{rfv}\nonumber
\Pi (T) =T^{n+1}f_n ({E_c\over T})
\sum\limits_{k=0}^{\infty}{a_k \over 2^k} \{
\sum\limits_{q=0}^{k} {k!\over (k-q)! q! }({T\over T_-})^q \} &\\
=T^{n+1}f_n ({E_c\over T})\psi (u_0) \;. &
\end{eqnarray}
The above approximation is useful when ${T\over T_-}\to 0$, i.e., 
the decay heavy meson mass is very close to the parent hybrid 
meson mass. We shall use the exact expression (\ref{edc}) in 
the numerical analysis below.

\section{Numerical results and discussions}
\label{sec5}
\setcounter{equation}{0}

We now turn to the numerical evaluation of the sum rules for the coupling
constants after the continuum is subtracted carefully.

The variation of the coupling constants $g_i$ with the Borel parameter 
$T$ and $E_c$ is presented in FIG. 6-12. The curves correspond to 
$E_c =1.4, 1.5, 1.6$GeV respectively. Stability develops for these sum 
rules starting from $0.8$ GeV. 
Numerically we have 
\begin{eqnarray}
\label{nu1}
 &&g_1  f_{-,{1\over 2} } f_H =-(0.12\pm 0.02)\mbox{GeV}^4\;,\\
 &&g_2  f_{-,{1\over 2} } f_H =(0.078\pm 0.016)\mbox{GeV}^4\;,\\
 &&g_3  f_{+,{1\over 2} } f_H =(0.11\pm 0.04)\mbox{GeV}^5\;,\\
 &&g_4  f_{+,{1\over 2} } f_H =(0.046\pm 0.01)\mbox{GeV}^3\;,\\
 &&g_5  f_{+,{3\over 2} } f_H =(0.33\pm 0.08)\mbox{GeV}^6\;,\\
 &&g_6  f_{+,{3\over 2} } f_H =(0.026\pm 0.006)\mbox{GeV}^4\;,\\
 &&g_7  f_{+,{3\over 2} } f_H =-(0.018\pm 0.004)\mbox{GeV}^4\;,
\end{eqnarray}
where the errors refers to the variations with $T$ and 
$E_c$ in this region. And the central value corresponds 
to $T=1.1$ GeV, $\beta =1$, and $E_c =1.5$GeV. 
Note the sum rules for $g_1, g_4, g_6$ changes less than $10\%$ 
when either $T$ or $E_c$ varies in the working region.
Variation with $T$ is about $20\%$ for those sum rules for 
$g_2, g_3, g_7$. Only the sum rule for $g_5$ has a rather strong 
dependence on the continuum threshold $E_c$. It changes about 
$20\%$ with $E_c=(1.5\pm 0.1)$ GeV. 

With the values of $f_i$ given in the previous sections we get
\begin{eqnarray}
\label{num1}
 &&g_1  =-(1.4\pm 0.3)\mbox{GeV}^{-1}\;,\\
 &&g_2  =(0.85\pm 0.2)\mbox{GeV}^{-1}\;,\\
 &&g_3  =(1.3\pm 0.3)\;,\\
 &&g_4  =(0.5\pm 0.2)\mbox{GeV}^{-2}\;,\\
 &&g_5  =(5.0\pm 1.0)\;,\\
 &&g_6  =(0.4\pm 0.1)\mbox{GeV}^{-2}\;,\\
 &&g_7  =-(0.3\pm 0.05)\mbox{GeV}^{-2}\;.
\end{eqnarray}

With these coupling constants we can calculate the decay widths of 
heavy hybrid mesons. The leading order binding energy of
$H_1$ meson is $\Lambda_H =1.6$ GeV, which is not small compared
with the bottom quark mass $m_b =4.7$GeV. We shall take into account 
of the corrections due to the finite $m_b$ partly. 
In full QCD, we introduce the decay constant of the hybrid meson $H_1$ as:
\begin{equation}\label{over-1}
\langle 0 | J_\mu (0) | H_1\rangle =m_H F_H \epsilon^1_\mu \; ,
\end{equation}
where $F_H={f_H\over \sqrt{m_b}}$ as $m_b\to \infty$.
The amplitude of the decay process $H_1\to B \pi$ in full QCD is 
\begin{equation}\label{am-1}
 M(H_1\to B \pi  )= \epsilon_1^\mu q_\mu \sqrt{m_H m_B}h_1 \;.
\end{equation}
In the limit $m_b\to \infty$ the decay constant $h_1$ in full QCD 
is equal to the decay constant $g_1$ defined in HQET.
The decay width formulas in the leading order of HQET are not
useful in the present case. We include the finite $m_b$ correction and
use the following decay width formulas.
\begin{eqnarray}
\label{widths}
&&\Gamma(H_1\to B\pi )= {1\over 16\pi} ({m_B\over m_H})
({m_b^2\over m_Bm_H})^2 g_1^2|\vec q|^3
\;,\nonumber\\
&&\Gamma(H_1\to B^*\pi )= {1\over 8\pi} ({m_{B^*}\over m_H})
({m_b^2\over m_{B^*}m_H})^2 g_2^2|\vec q|^3\;,\nonumber\\
&&\Gamma(H_1\to B'_1\pi )= {1\over 16\pi} ({m_{B'_1}\over m_H})
({m_b^2\over m_{B'_1}m_H})^2 
(3g_3^2+4g_3g_4|\vec q|^2 +2g_4 |\vec q|^4) |\vec q|\;,\nonumber\\
&&\Gamma(H_1\to B_1\pi )=
{1\over 16\pi} ({m_{B_1}\over m_H})
({m_b^2\over m_{B_1}m_H})^2 
(3g_5^2+4g_5 g_6|\vec q|^2 +2g_6 |\vec q|^4) |\vec q|\;,\nonumber\\
&&\Gamma(H_1\to B_2^*\pi)= {1\over 4\pi} ({m_{B_2^*}\over m_H})
({m_b^2\over m_{B_2^*}m_H})^2 g_7^2|\vec q|^5  \;,
\end{eqnarray}
where $|\vec q|=\sqrt{(m_H^2-(m_Q+m_\pi )^2)(m_H^2-(m_Q-m_\pi )^2)}/2m_H$, 
$m_H$, $m_Q$ is the hybrid and heavy meson mass.
Note summation over charged and neutral pion final states has been performed.
In the calculation we use $m_B=5.28$ GeV, $m_{B^*}=5.33$ GeV, 
$m_{B'_1}=m_B+0.37=5.7$ GeV, $m_{B_1}=m_{B_2^*}=5.8$ GeV, 
$m_H=m_b+\Lambda_H =6.3$ GeV with $m_b=4.7$ GeV. For the heavy meson
with strangeness, we use $m_{B_s}=m_B+0.09=5.37$ GeV etc.

We employ the $SU_f(3)$ flavor symmetry to relate the coupling constants
for the decay processes $H_1\to B_s K$, $H_1\to B\eta$ etc. 
For example, $g_{H_1\to B_s K}=g_1, g_{H_1\to B\eta }={g_1\over \sqrt{6}}$.
We collect the numerical results of the decay widths for the 
different channels in TABLE VI.

Summing all the two-body strong decay channels
with one Goldstone boson in the final state, we estimate the
total decay width of the lowest lying hybrid meson 
to be around $300$ MeV. The dominant decay mode is $H_1\to B_1 \pi $. 
Its branching ratio is about $80\%$. Now we know $B_1$ is a
narrow resonance with a width of $\sim 20$ MeV. So this mode
can be used to detect the possible existence of $H_1$ experimentally. 
The same characteristic decay modes for the light hybrid mesons have 
been predicted in the flux tube model \cite{flux}. 

In summary we have calculated the binding energy of the heavy hybrid
mesons in the leading order of HQET. With the help of the 
light cone QCD sum rules we have extracted the pionic couplings
between the hybrid $H_1$ and lowest three heavy meson doublets.
Invoking the flavor $SU_f(3)$ symmetry we have estimated the strong
two-body decay widths of $H_1$ with one Goldstone boson in the 
final state. Our calculation yields very characteristic decay 
modes of $H_1$ and confirms the earlier predictions on the 
particular decay modes for the light hybrid mesons from the 
flux tube model. The mixing between $\eta_8$ and $\eta_1$ 
and the possible contribution due to the QCD anomaly will be 
topics of fututre work.

\vspace{0.8cm} {\it Acknowledgments:\/}This project was supported by
the Natural Science Foundation of China.

\newpage

TABLE I. The coefficients in the expressions for the spectral density
$\rho (\epsilon )$.
\vskip 0.5cm
\begin{tabular}{|c|c|c|c|c|}
\hline
 & $1^{-+}$ & $0^{++}$ &$1^{+-}$  & $0^{--}$   \\
\hline
$c_1$ &$1$&$3$&$-1$&$-3$  \\ 
$c_2$ &$1$&$3$&$-1$&$-3$  \\ 
$c_3$ &$1$&$-3$&$1$&$-3$  \\ 
$c_4$ &$1$&$-3$&$1$&$-3$  \\ 
$c_5$ &$1$&$-1$&$-1$&$1$  \\ 
\hline
\end{tabular}
\vskip 1.0 true cm

TABLE II. The values of $\Lambda, f, s_0$ for the nonstrange 
heavy hybrid meson $(Q\bar q g )$. $\Lambda, s_0$ is unit of GeV, 
$F$ is in unit of GeV$^{7\over 2}$.
\vskip 0.5cm
\begin{tabular}{|c|c|c|c|c|}
\hline
 & $1^{-+}$ & $0^{++}$ &$1^{+-}$  & $0^{--}$   \\
\hline
$\Lambda$ &$1.6$& $2.17$&$1.66$&$3.0$  \\ 
$f$ &$0.23$& $0.425$&$0.162$&$0.98$  \\ 
$s_0$ &$2.0$& $2.5$&$1.9$&$3.4$  \\ 
\hline
\end{tabular}
\vskip 1.0 true cm

TABLE III. The ratio between various condensates and perturbative 
term after the continuum subtraction for the nonstrange heavy hybrid meson.
\vskip 0.5cm
\begin{tabular}{|c|c|c|c|c|}
\hline
 & $1^{-+}$ & $0^{++}$ &$1^{+-}$  & $0^{--}$   \\
\hline
$<\bar q q>$ &$0.22$& $0.34$&$-0.26$&$-0.14$  \\ 
$<g_s^2G^2>$ &$0.16$& $-0.23$&$0.19$&$-0.08$  \\ 
$<g_s^3G^3>$ &$0.004$& $-0.004$&$0.005$&$-0.0006$  \\ 
$<\bar q q><g_s^2G^2>$ &$0.02$& $-0.004$&$-0.026$&$0.0006$  \\ 
\hline
\end{tabular}
\vskip 1.0 true cm

\newpage

TABLE IV. The values of $\Lambda, f, s_0$ for the strange 
heavy hybrid meson $(Q\bar q g )$. 
\vskip 0.5cm
\begin{tabular}{|c|c|c|c|c|}
\hline
 & $1^{-+}$ & $0^{++}$ &$1^{+-}$  & $0^{--}$   \\
\hline
$\Lambda$ &$1.73$& $2.28$&$1.76$&$3.15$  \\ 
$f$ &$0.28$& $0.538$&$0.2$&$1.04$  \\ 
$s_0$ &$2.13$& $.63$&$2.03$&$3.53$  \\ 
\hline
\end{tabular}
\vskip 1.0 true cm

TABLE V. The ratio between various condensates (including strange quark 
mass correction) and perturbative term for the strange heavy hybrid meson.
\vskip 0.5cm
\begin{tabular}{|c|c|c|c|c|}
\hline
 & $1^{-+}$ & $0^{++}$ &$1^{+-}$  & $0^{--}$   \\
\hline
$<\bar s s>$ &$0.15$& $0.23$&$-0.17$&$-0.1$  \\ 
$<g_s^2G^2>$ &$0.13$& $-0.19$&$0.15$&$-0.07$  \\ 
$<g_s^3G^3>$ &$0.003$& $-0.003$&$0.004$&$-0.0005$  \\ 
$<\bar s s><g_s^2G^2>$ &$0.01$& $-0.003$&$-0.014$&$0.0004$  \\ 
$m_s$ & $0.08$&$0.20$&$-0.09$&$-0.15$  \\ 
\hline
\end{tabular}
\vskip 1.0 true cm

TABLE VI. The coefficients $a_k$ when the PWFs are expnaded into 
polynomials of $(1-u)$. They are in unit of $\delta^2$ for the 
twist four PWFs.
\vskip 0.5cm
\begin{tabular}{|c|c|c|c|c|c|}
\hline
 & $a_1$ & $a_2$ &$a_3$  & $a_4$ &$a_5$  \\ 
 \hline
$\Phi_{3\pi} (u)$ & $$ & $$ &$-120$  & $150$ &$$  \\ 
$\Phi_\bot (u)$& $$ & $$ &${10\over 3}$  & $-{25\over 6}$ &$$  \\ 
$\Phi_\bot (u)-{\tilde \Phi}_\bot (u)
+{\tilde\Phi}_{\|}(u)$ & $$ & $$ &$-{20\over 3}$  & $5$ &$$  \\ 
$\Phi_{3\pi}^{''}(u)$ & $$ & $360$ &$-600$  & $$ &$$  \\ 
${\tilde\Phi}'_\bot (u)$ & $$ & $-10$ &$10$  & $$ &$$  \\ 
$\Phi_{3\pi}(u)$ & $$ & $$ &$$  & $30$ &$-30$  \\ 
$[u\Phi_{3\pi}(u)]^{'''}$& $-720$ & $3600$ &$-3600$  & $$ &$$  \\ 
$\{ u [\Phi_\bot (u)-{\Phi_{\|}(u) \over 2}]\}^{''}$& $20$ 
   & $-90$ &${250\over 3}$  & $$ &$$  \\ 
$[u\Phi_{3\pi}(u)]'$ & $$ & $$ &$-120$  & $300$ &$-180$  \\ 
$u[ 5\Phi_\bot (u) -2\Phi_{\|}(u)
-{\tilde\Phi}_\bot (u)+{\tilde\Phi}_{\|}(u)]$
 & $$ & $$ &${20\over 3}$  & $-{55\over 3}$ &${35\over 3}$  \\ 
$u [\Phi_\bot (u)-{\tilde\Phi}_\bot (u)+{\tilde\Phi}_{\|}(u)]$
 & $$ & $$ &$-{20\over 3}$  & ${35\over 3}$ &$-5$  \\ 
\hline
\end{tabular}
\vskip 1.0 true cm

TABLE VII. The decay widths of different two-body decay 
channels for the $1^{-+}$ hybrid meson $H_1$, where 
combinations of Goldstone bosons and heavy mesons yield 
different final states. The unit is MeV. The minus sign means
either such a decay mode is not allowed by the phase space
or the decay width is negligible.
\vskip 0.5cm
\begin{tabular}{|c|c|c|c|c|c|c|}
\hline
 & $B$ & $B^*$ &$B'_0$  & $B'_1$&$B_1$&$B_2^*$   \\
\hline
$\pi$ &$11$& $9$&$0$&$25$&$230$&$0.05$  \\ 
$\eta$ &$1$&$1$&$0$&$1.6$&-& -  \\ 
$K$ &$5$&$5$&$0$&$10$&-& -  \\ 
\hline
\end{tabular}
\vskip 1.0 true cm

\newpage
{\bf Figure Captions}
\vspace{2ex}
\begin{center}
\begin{minipage}{130mm}
{\sf FIG. 1.} \small{The relevant feynman diagrams for the derivation of 
the QCD sum rule (\ref{mass-1}). The broken solid line, broken curly line
and a broken solid line with a curly line attached in the middle
stands for the quark condensate, gluon condensate 
and quark gluon mixed condensate respectively. }
\end{minipage}
\end{center}
\begin{center}
\begin{minipage}{130mm}
{\sf FIG. 2} 
\small{The variations of $\Lambda$ with $T$ and $s_0$ for $H_1$. 
From top to bottom the curves correspond to $s_0 =2.1, 2.0, 1.9$ GeV.
$T$ is in unit of GeV.}
\end{minipage}
\end{center}
\begin{center}
\begin{minipage}{130mm}
{\sf FIG. 3} 
\small{The variations of $\Lambda$ with $T$ and $s_0 =3.5, 3.4, 3.3$ 
GeV for $H_4$.}
\end{minipage}
\end{center}
\begin{center}
\begin{minipage}{130mm}
{\sf FIG. 4} \small{The variation of the right and left hand side 
of Eq. (\ref{mass-1}) with $T$ is plotted as solid and dotted curves 
respectively for $H_1$ with the values of $\Lambda, f, s_0$ in TABLE II. }
\end{minipage}
\end{center}
\begin{center}
\begin{minipage}{130mm}
{\sf FIG. 5} \small{The variation of the right and left hand side 
of Eq. (\ref{mass-1}) with $T$ for $H_4$. }
\end{minipage}
\end{center}
\begin{center}
\begin{minipage}{130mm}
{\sf FIG. 6} \small{The dependence of $g_1 f_{-,{1\over 2}} f_H$ 
on the Borel parameter $T$ and the continuum threshold $E_c$ after
the factor ${1\over 4} 
e^{-( { \Lambda_{-,{1\over 2} } \over T_1}+{ \Lambda_H \over T_2} )}
$ is moved to the right hand side of Eq. (\ref{g1}). }
\end{minipage}
\end{center}
\begin{center}
\begin{minipage}{130mm}
{\sf FIG. 7} \small{The variation of $g_2 f_{-,{1\over 2}} f_H$  
with $T$ and $E_c$. }
\end{minipage}
\end{center}
\begin{center}
\begin{minipage}{130mm}
{\sf FIG. 8} \small{The variation of $g_3 f_{+,{1\over 2}} f_H$  
with $T$ and $E_c$. }
\end{minipage}
\end{center}
\begin{center}
\begin{minipage}{130mm}
{\sf FIG. 9} \small{The variation of $g_4 f_{+,{1\over 2}} f_H$  
with $T$ and $E_c$. }
\end{minipage}
\end{center}
\begin{center}
\begin{minipage}{130mm}
{\sf FIG. 10} \small{The variation of $g_5 f_{+,{3\over 2}} f_H$  
with $T$ and $E_c$. }
\end{minipage}
\end{center}
\begin{center}
\begin{minipage}{130mm}
{\sf FIG. 11} \small{The variation of $g_6 f_{+,{3\over 2}} f_H$  
with $T$ and $E_c$. }
\end{minipage}
\end{center}
\begin{center}
\begin{minipage}{130mm}
{\sf FIG. 12} \small{The variation of $g_7 f_{+,{3\over 2}} f_H$  
with $T$ and $E_c$. }
\end{minipage}
\end{center}


\bigskip
\vspace{1.cm}

\begin{thebibliography}{99}
\bibitem{e852}D.R.Thompson et al., Phys. Rev. Lett. {\bf 79} (1997) 1630.

\bibitem{cb}A.Abele et al., Phys. Lett. B 423 (1998) 175.

\bibitem{ves}G.M.Beladidze et al., Phys. Lett. B {\bf 313} (1993) 276.

\bibitem{balitsky}
I.I.Balitsky, D.I.Dyakanov and A.V.Yung, Z. Phys. C 33 (1986) 265.

\bibitem{latorre}
J.I.Latorre, P.Pascual and S.Narison, Z. Phys. C 34 (1987) 347.

\bibitem{deviron}J.Govaerts et al., Nucl. Phys. B 248 (1984) 1.

\bibitem{svz}M.A. Shifman, A.I. Vainshtein and V.I. Zakharov, 
Nucl. Phys. {\bf B 147} (1979) 385. 

\bibitem{deviron-heavy}
J.Govaerts et al., Nucl. Phys. B 258 (1985) 215; 
Nucl. Phys. B 262 (1985) 575; Nucl. Phys. B 284 (1987) 674.

\bibitem{deviron-decay}J.Govaerts et al., Phys. Rev. Lett 53 (1984) 2207.

\bibitem{zhu-bbg}Shi-Lin Zhu, hep-ph/9812467.

\bibitem{gluon-mass}D. Horn and J. Mandula, Phys. Rev. D 17, 898 (1978).

\bibitem{flux-mass}N. Isgur and J. Paton, Phys. Lett. 124B. 247 (1983);
Phys. Rev. D 31, 2910 (1985); T. Barnes, F. Close and E. S. Swanson, 
Phys. Rev. D 52, 5242 (1995).

\bibitem{mit}T. Barnes, Nucl. Phys. B 158, 171 (1979);
F. de Viron and J. Weyers, Nucl. Phys. B 185, 391 (1981);
T. Barnes and F. E. Close, Phys. Lett. 116 B, 365 (1982).

\bibitem{lattice}S. Perantonis and C. Michael, Nucl. Phys. B 347, 854 (1990);
C. Bernard et al., hep-lat/9809087; T. Manke et al., hep-lat/9812017;
G. S. Bali, hep-lat/9901023.

\bibitem{gluon-decay}M. Tanimoto, Phys. Lett. B 116, 198 (1982); 
Phys. Rev. D 27, 2648 (1978).

\bibitem{flux}N. Isgur, R. Kokoski and J. Paton, Phys. Rev. Lett. 54, 869 (1985);
F. E. Close and P. R. Page, Nucl. Phys. B 443, 233 (1995); 
Phys. Rev. D 52, 1706 (1995); P. R. Page, E. Swanson and 
A. P. Szczepaniak, Phys. Rev. D 59, 034016 (1999).

\bibitem{hqet}B. Grinstein, Nucl. Phys. {\bf B339} (1990) 253; 
E.Eichten and B. Hill, Phys. Lett. {\bf B234} (1990) 511; 
A. F. Falk et al., Nucl. Phys. {\bf B343} (1990) 1; 
F. Hussain et al., Phys. Lett. {\bf B249} (1990) 295;
H. Georgi, Phys. Lett. {\bf B240} (1990) 447. 

\bibitem{bely-89}I. I. Balitsky et al., Nucl. Phys. B 312 (1989) 509;
V. M. Braun and I. E. Filyanov, Z. Phys. C {\bf 44} (1989) 157.

\bibitem{bely-95}V. M. Belyaev et al., Phys. Rev. D 51 (1995) 6177.

\bibitem{zhu-pion}
Shi-Lin Zhu and Yuan-Ben Dai, Phys. Rev. D 58 (1998) 094033; 
Phys. Lett. B {\bf 429} (1998) 72; Yuan-Ben Dai and Shi-Lin Zhu, 
Euro. Phys. J. C 6 (1999) 307; Phys. Rev. D 58 (1998) 074009.

\bibitem{zhu-radiat}Shi-Lin Zhu and Yuan-Ben Dai, hep-ph/9810243.

\bibitem{zhit}A. R. Zhitnitsky, Sov. J. Nucl. Phys. {\bf 41} (1985) 1035.

\bibitem{huang}Y. B. Dai, C. S. Huang, M. Q. Huang and C. Liu, Phys. Lett. 
{\bf B390} (1997) 350.

\bibitem{neubert}E. Bagan, P. Ball, V. M. Braun and H. G. Dosch, Phys.
Lett. {\bf B278} (1992) 457; M. Neubert, Phys. Rev. {\bf D 45} (1992) 2451;
D. J. Broadhurst and A. G. Grozin, Phys. Lett. {\bf B274} (1992) 421.

\bibitem{zhu-duality}Shi-Lin Zhu and Yuan-Ben Dai, hep-ph/9811449.

\bibitem{f-omega}V. L. Chernyak and A. R. Zhitnitsky, 
Phys. Rep. {\bf 112}, 173(1984).

\bibitem{delta2}V. A. Novikov, M. A. Shifman, A. I. Vainshtein and V. I. Zakharov, 
Nucl. Phys. {\bf B 237}, 525(1984).

\bibitem{epsilon}V. M. Braun 
and I. B. Filyanov, Z. Phys. C {\bf 48}, 239(1990).

\end{thebibliography}
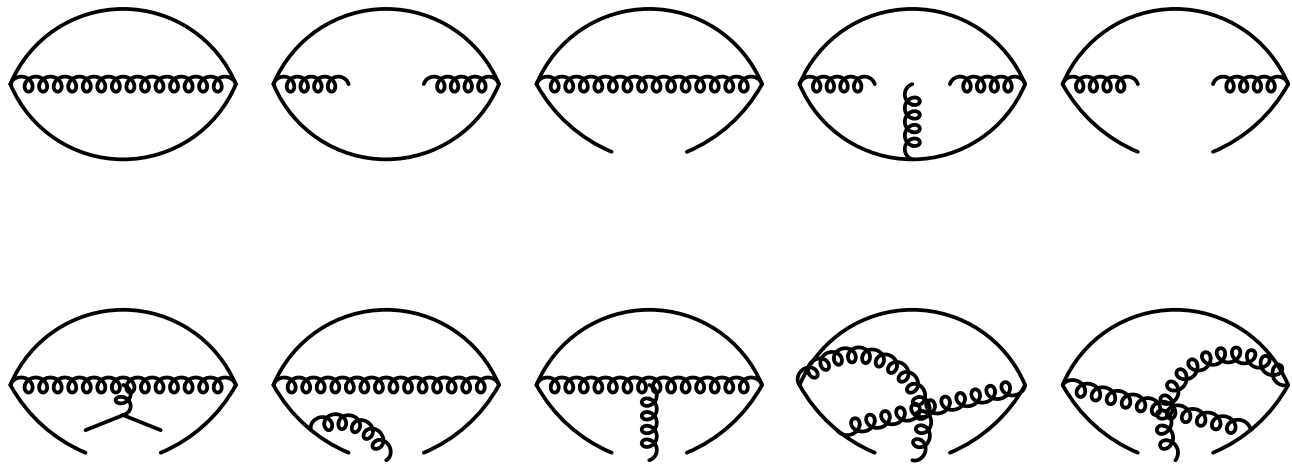
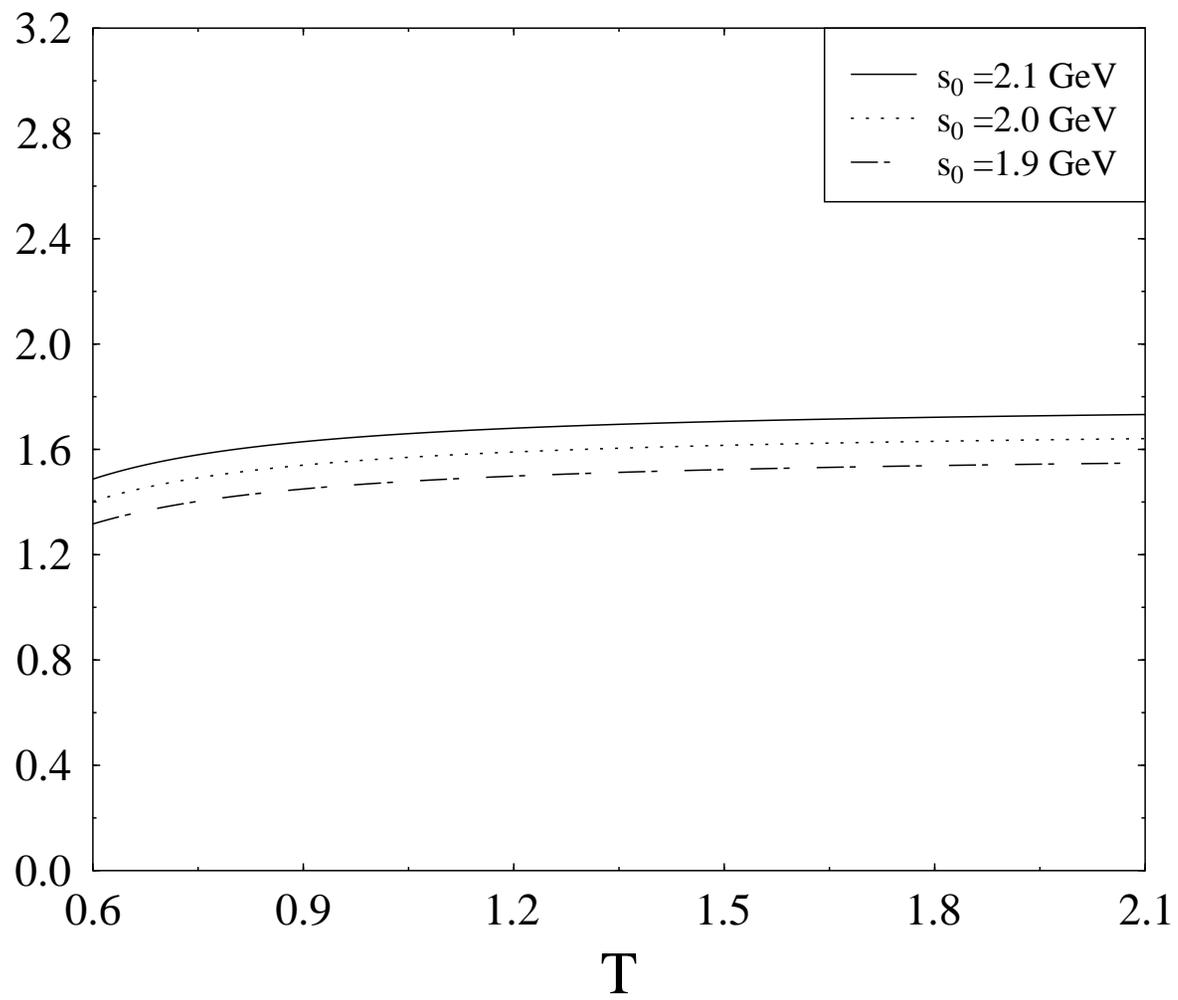
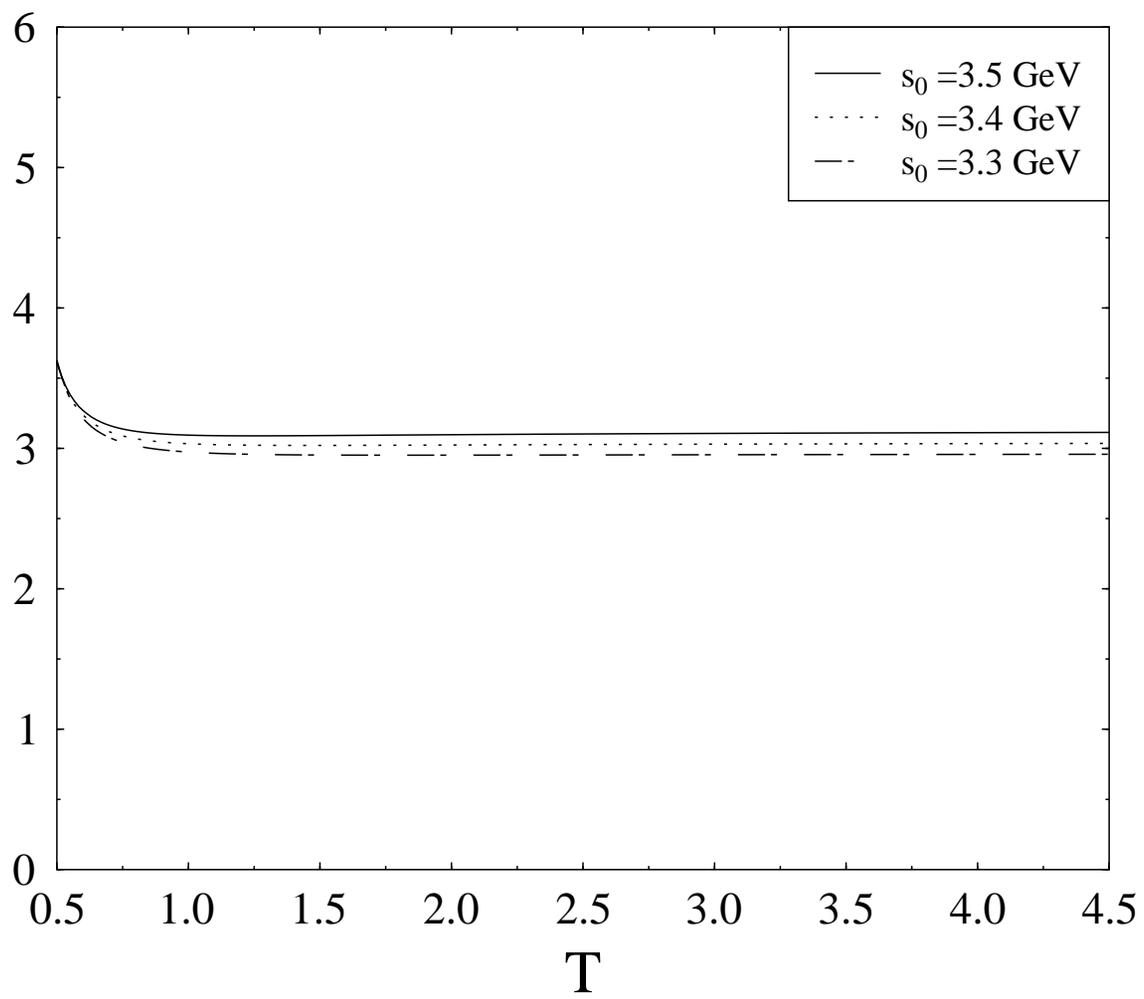
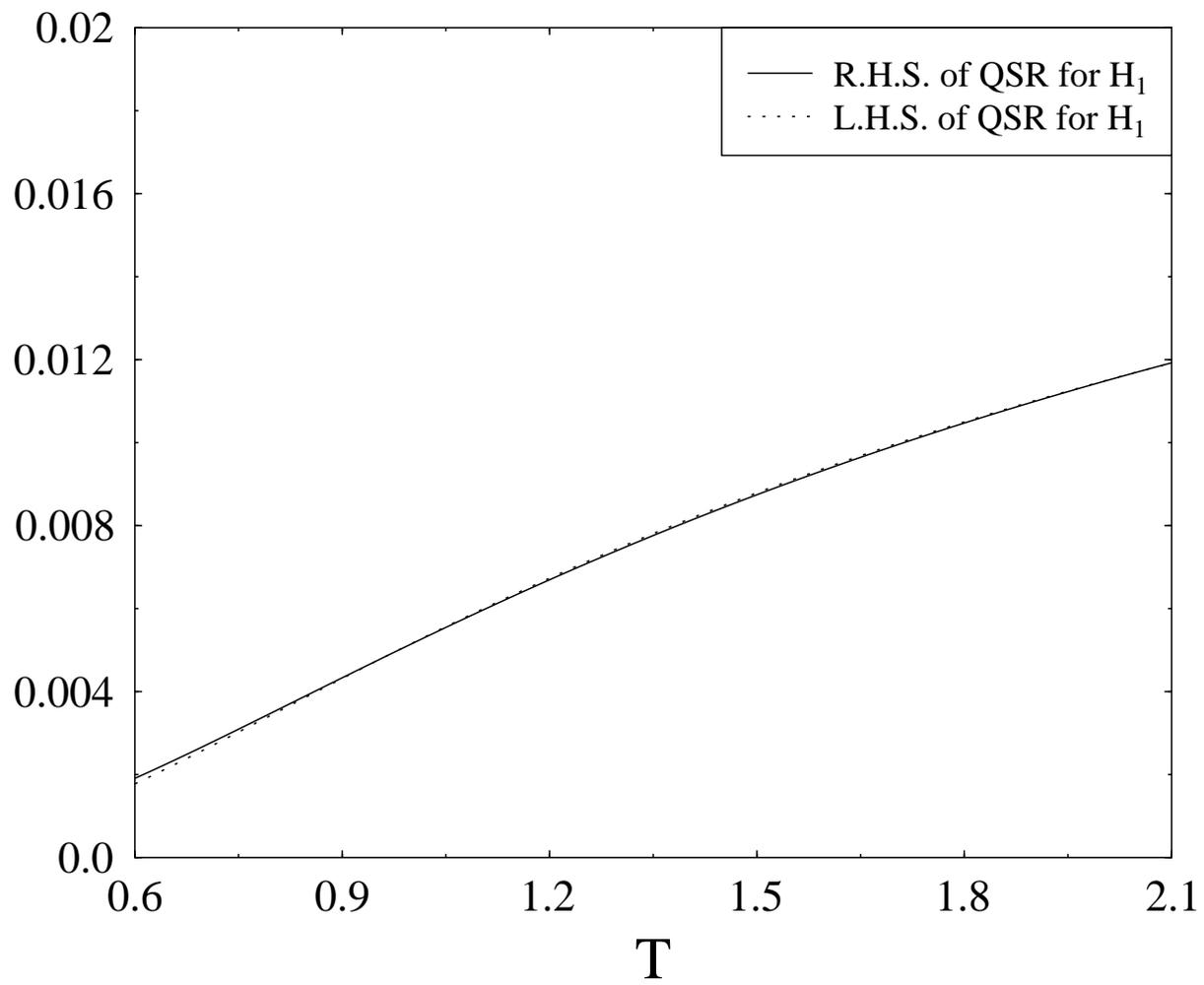
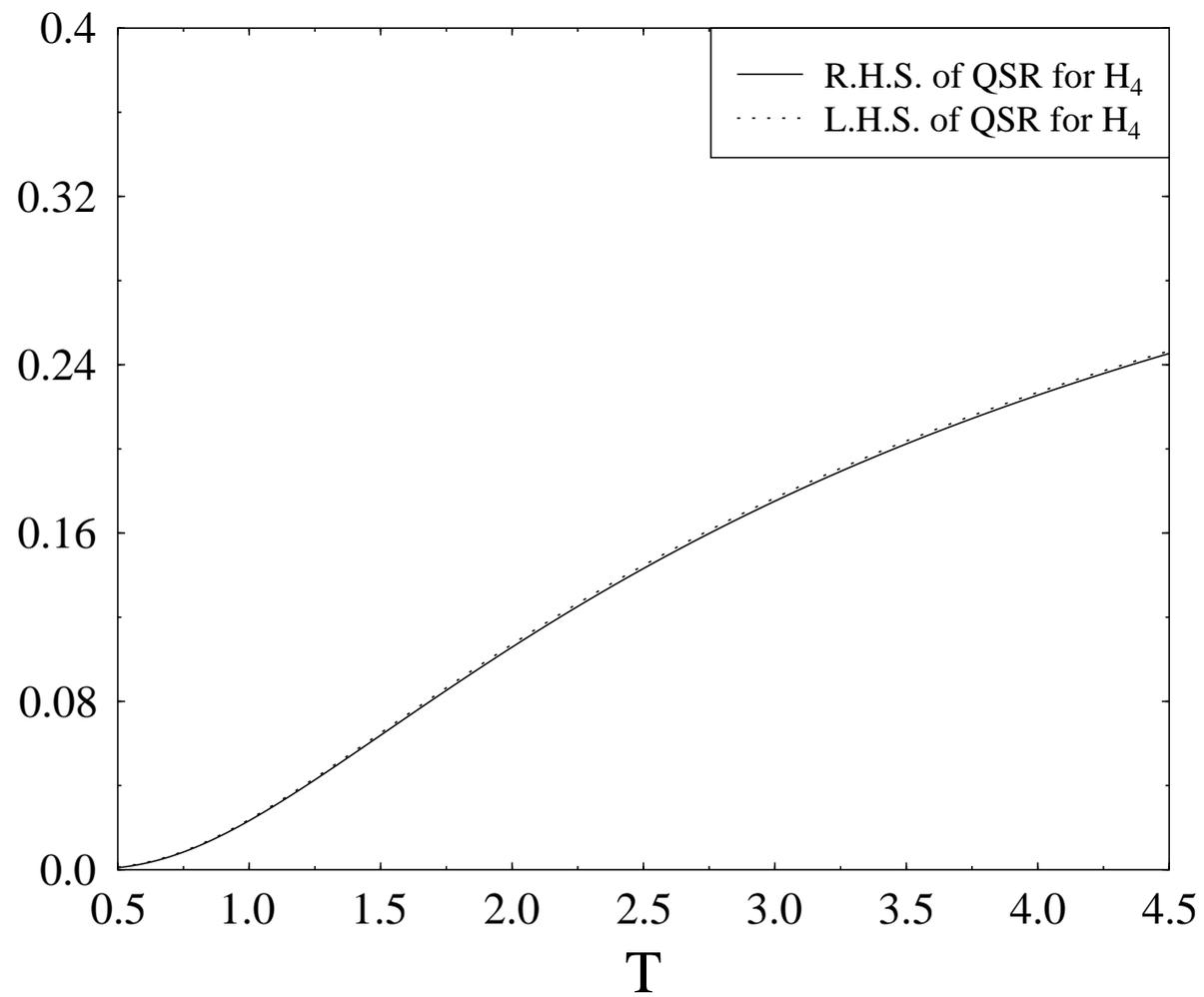
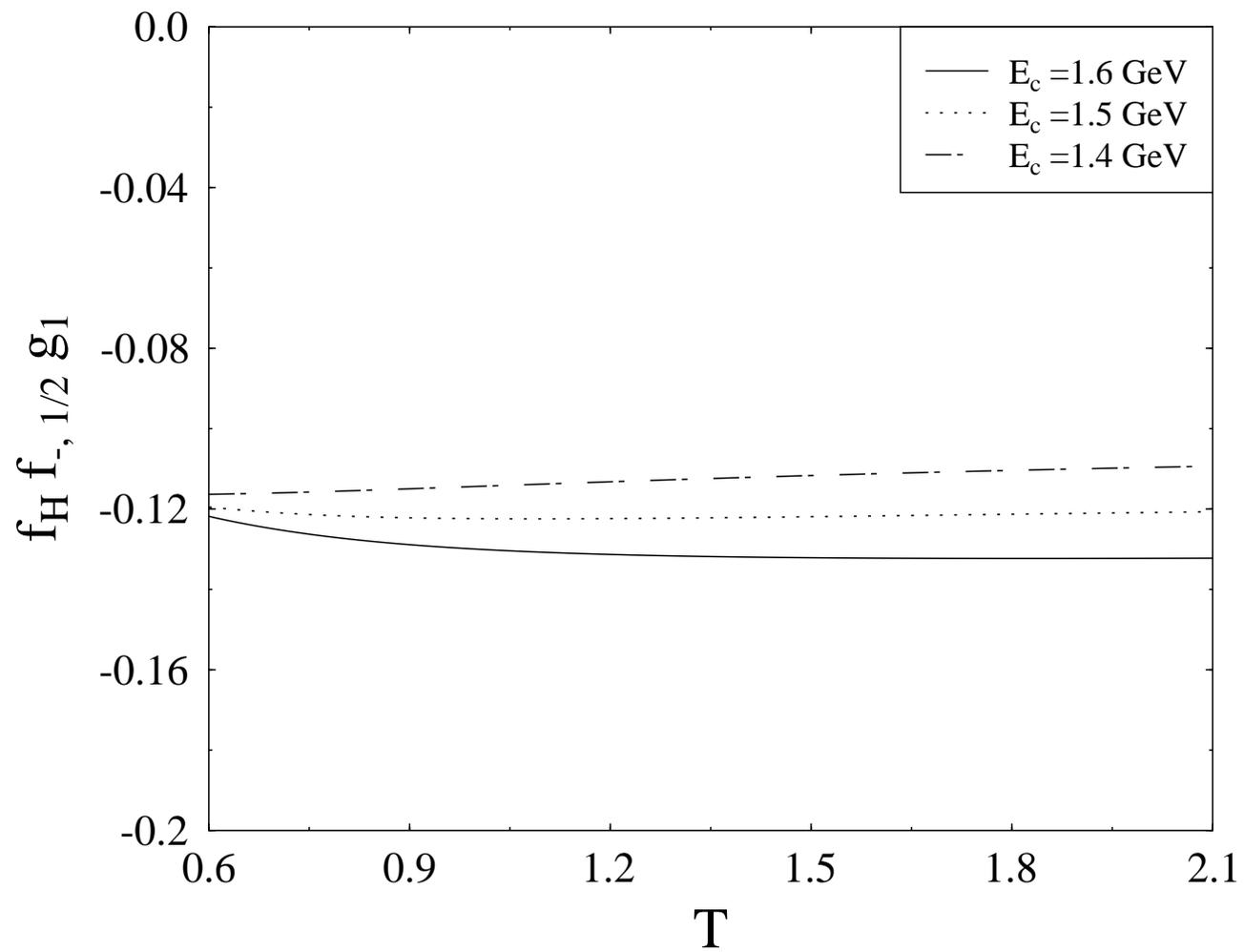
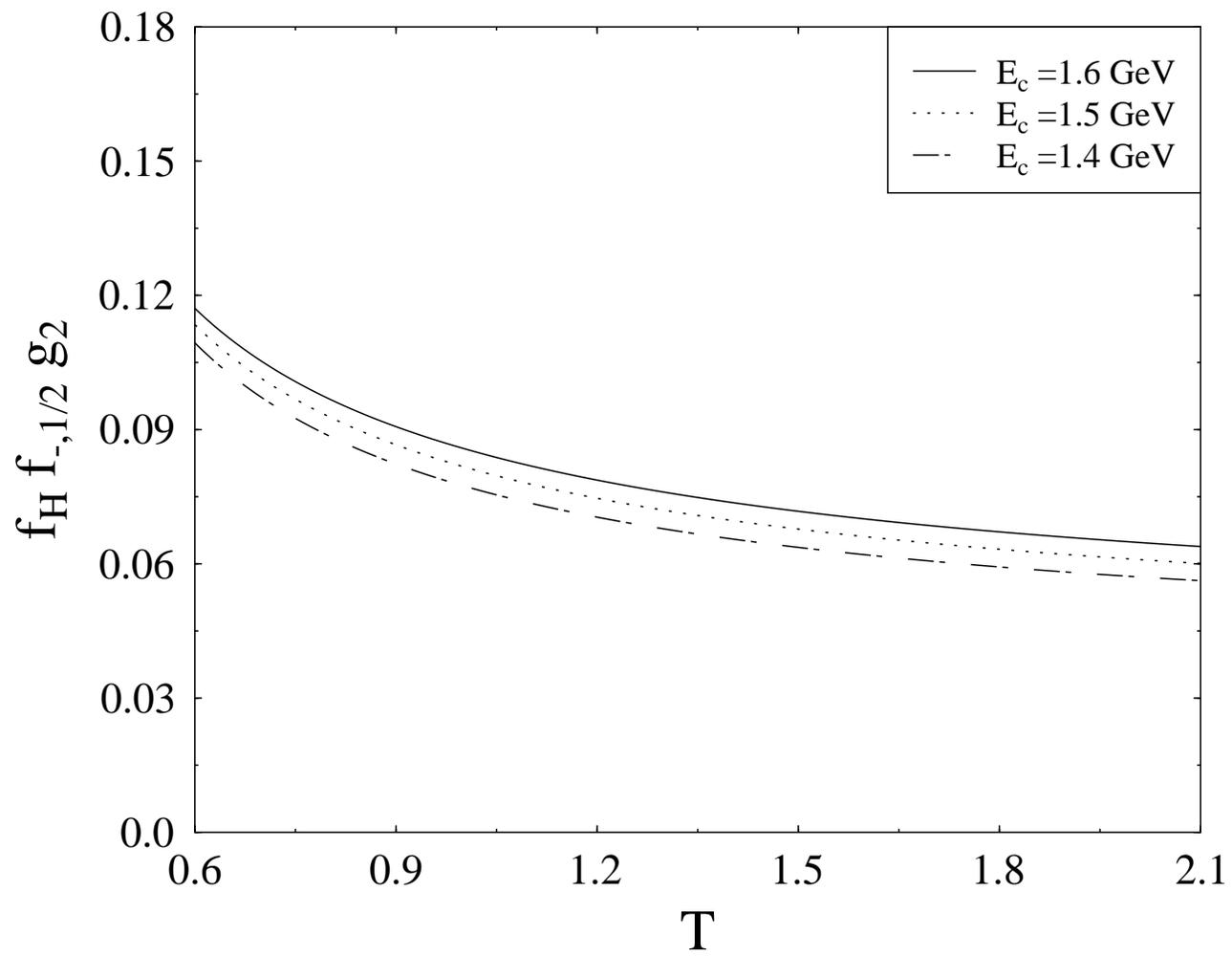
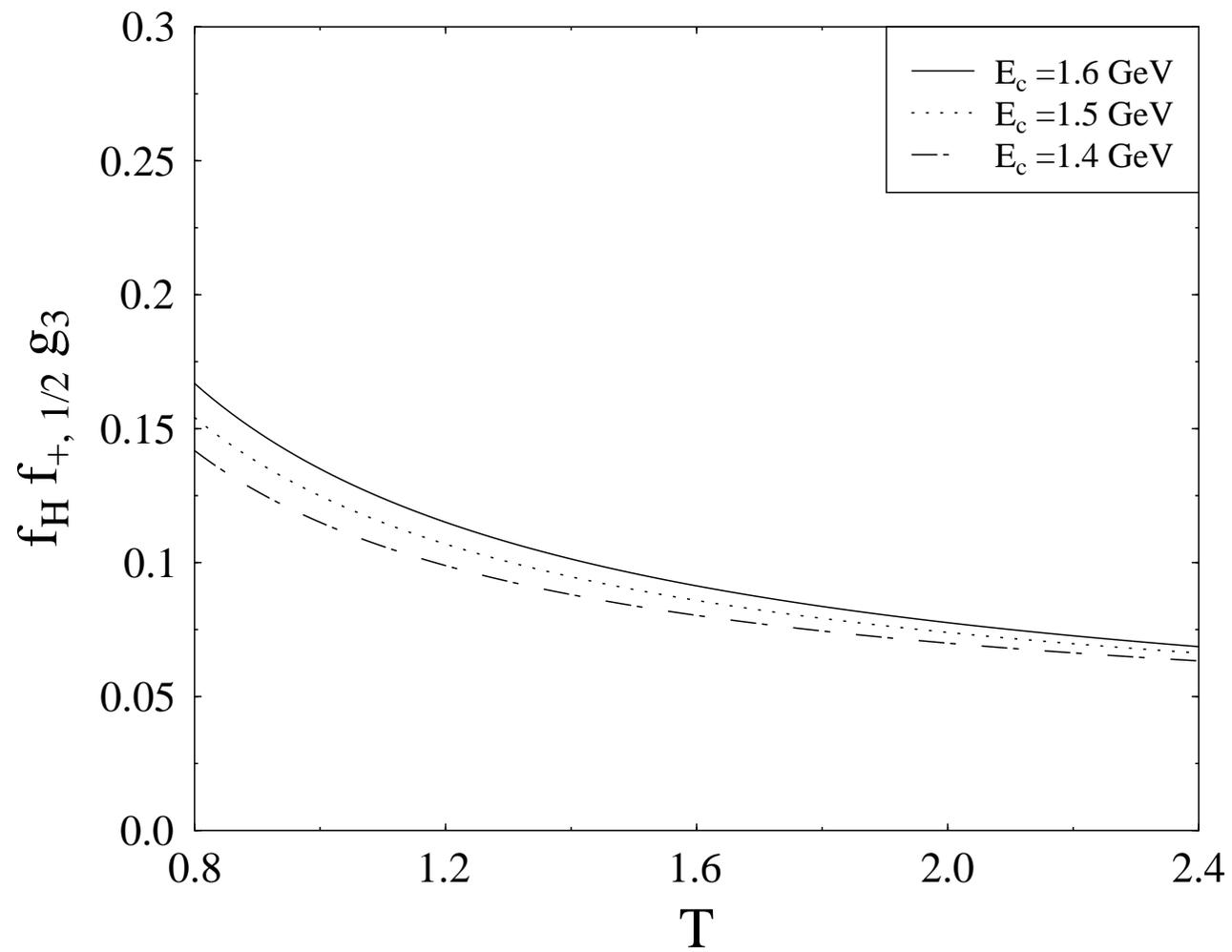
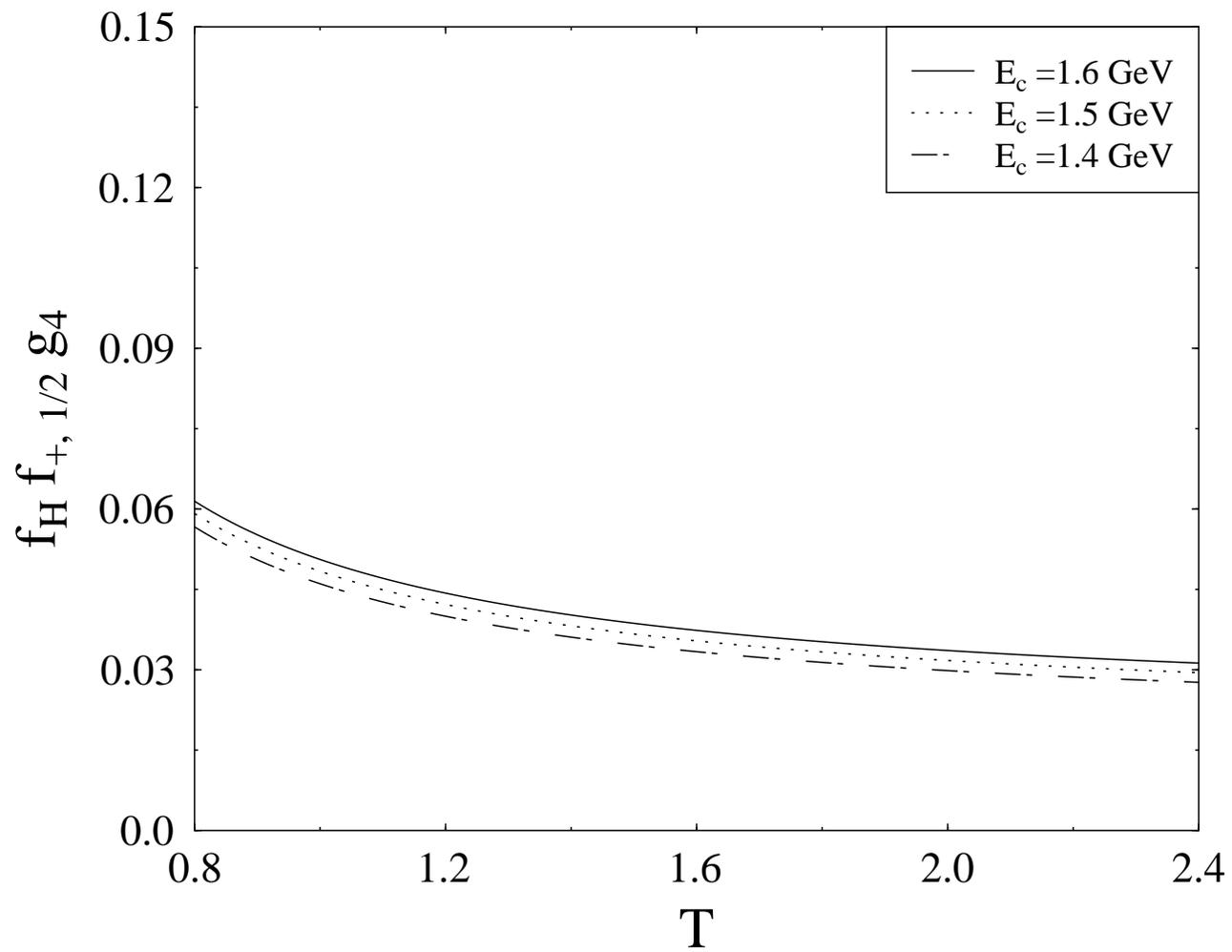
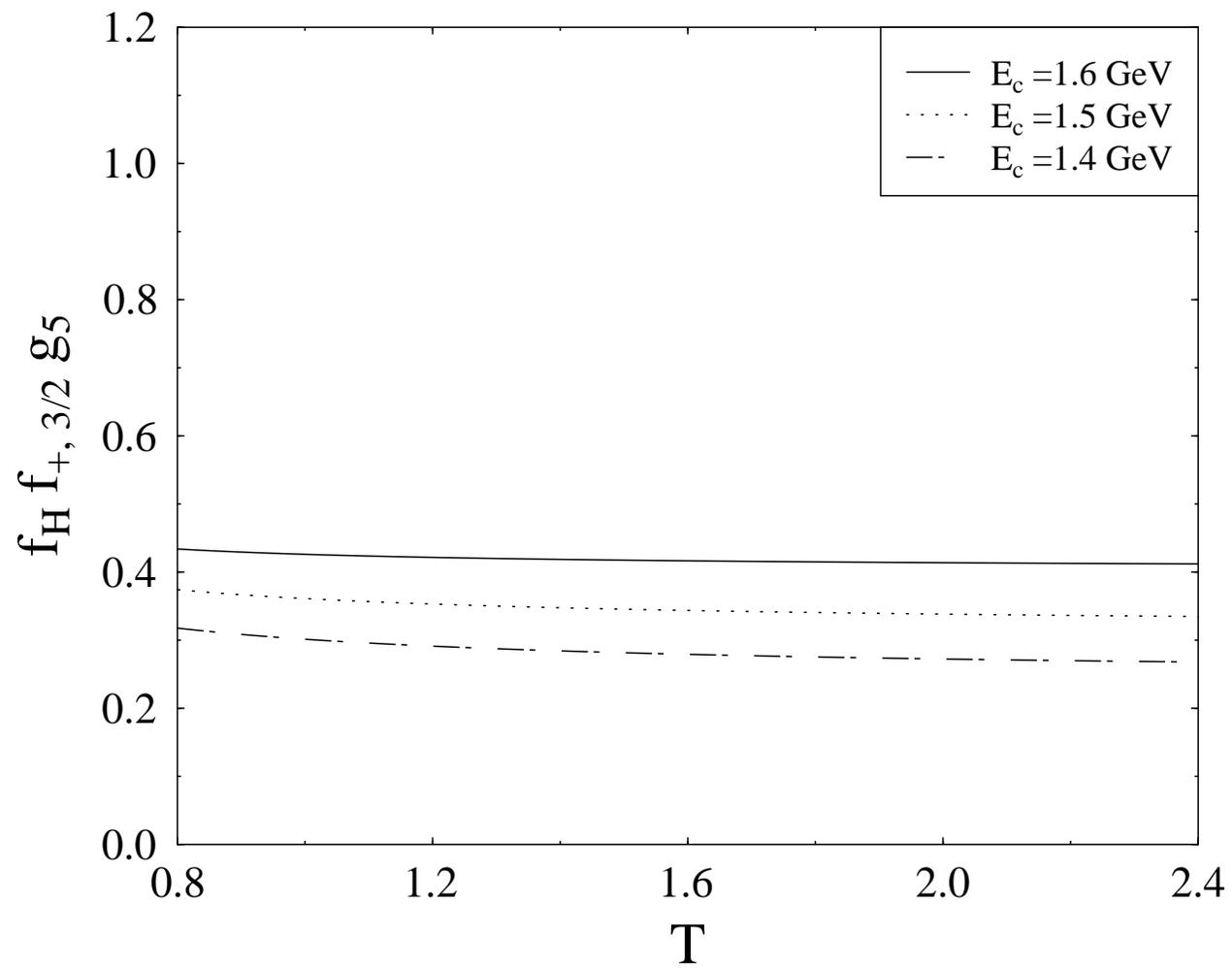
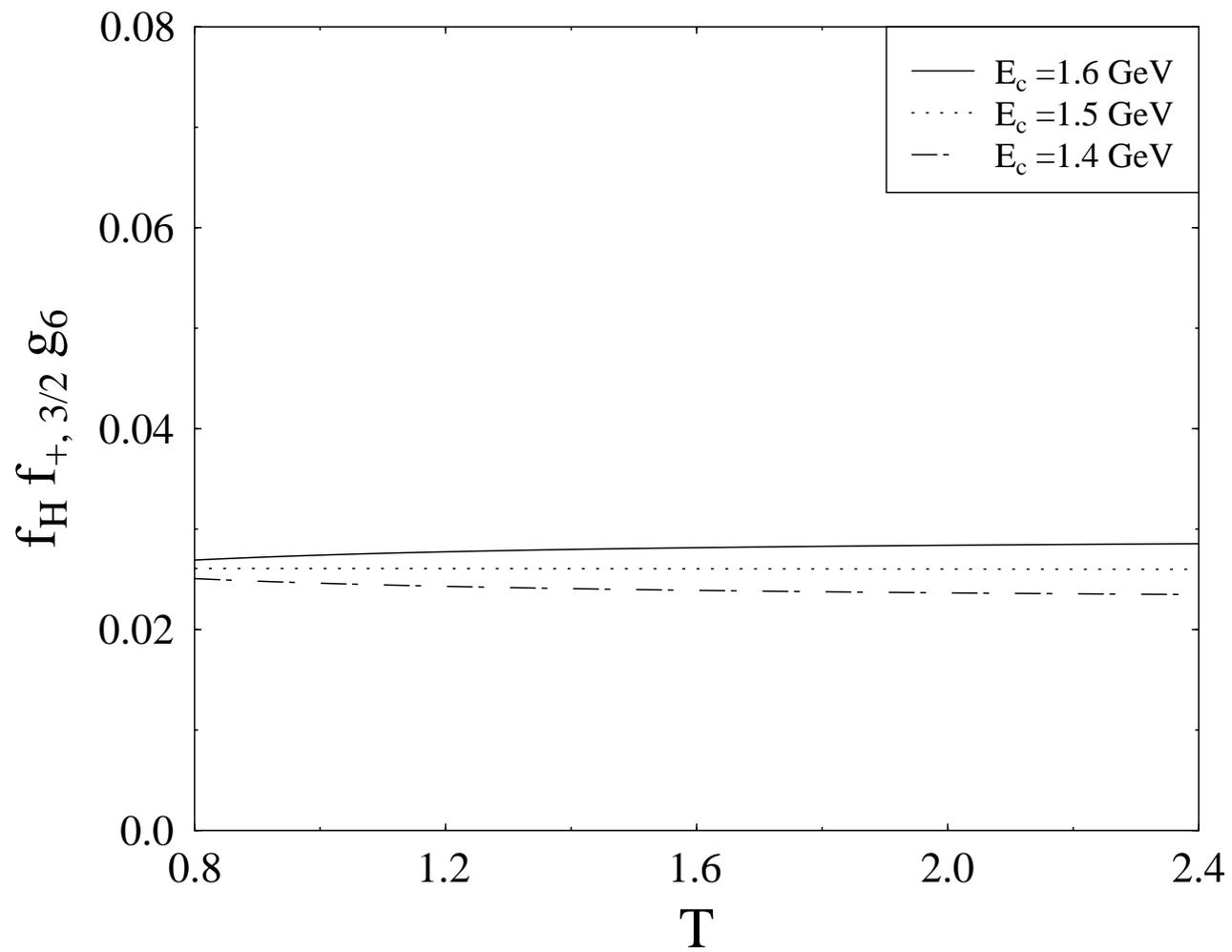
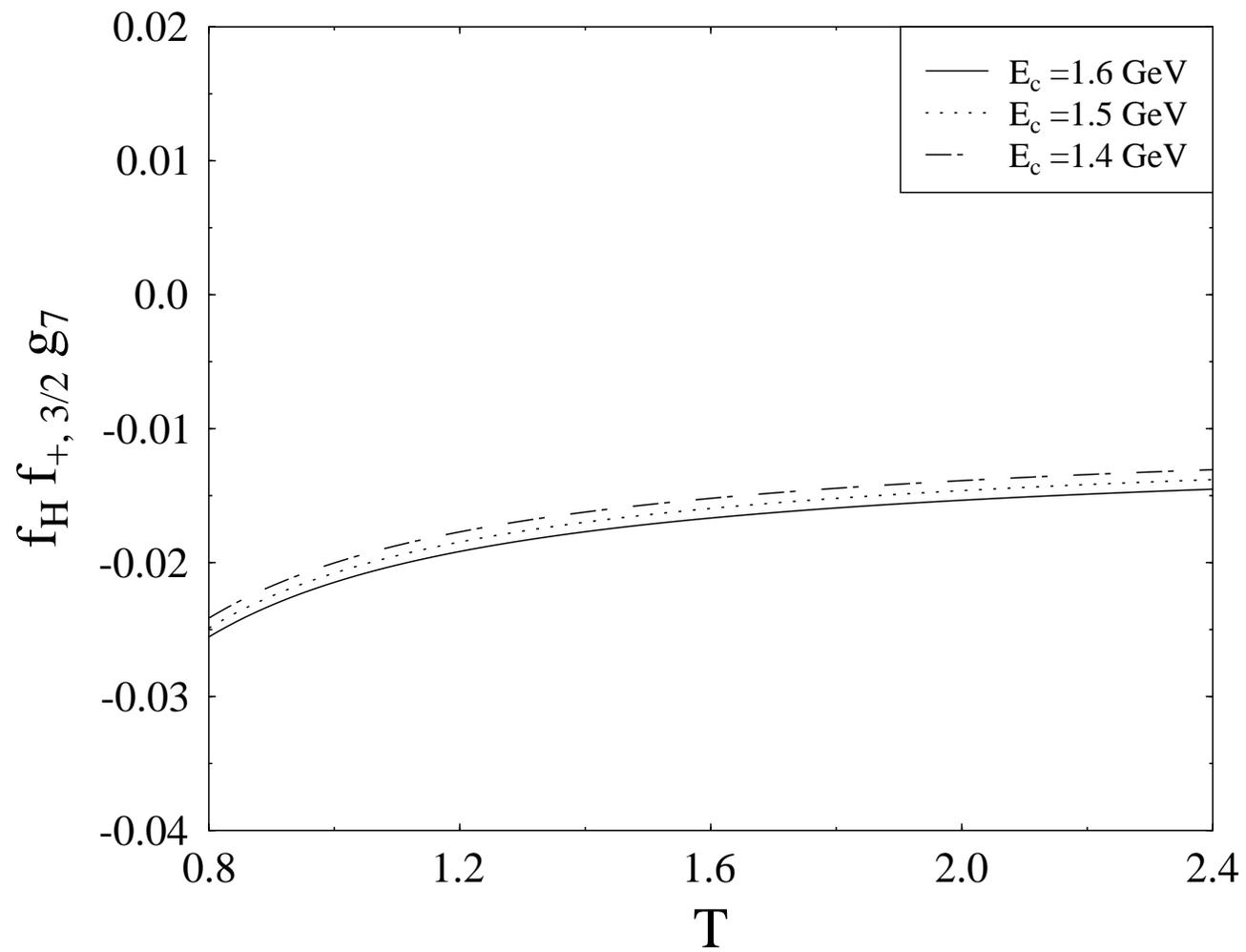
\end{document}